\documentclass[fleqn,usenatbib]{mnras}
\usepackage{newtxtext,newtxmath}
\usepackage{ulem}
\newcommand\redsout{\bgroup\markoverwith{\textcolor{red}{\rule[0.5ex]{2pt}{1.5pt}}}\ULon}

\usepackage[T1]{fontenc}
\usepackage{xcolor}
\usepackage{verbatim}
\DeclareRobustCommand{\VAN}[3]{#2}
\let\VANthebibliography\thebibliography
\def\thebibliography{\DeclareRobustCommand{\VAN}[3]{##3}\VANthebibliography}

\usepackage{graphicx}	
\usepackage{amsmath}	

\title[Fountain-driven corona accretion in NGC~2403]{Fountain-driven gas accretion feeding star formation over the disc of NGC~2403}

\author[A. Li et al.]{
Anqi Li$^{1}$\thanks{E-mail: li@astro.rug.nl},
Filippo Fraternali$^{1}$,
Antonino Marasco$^{2,3}$,
Scott C. Trager$^{1}$,
Gabriele Pezzulli$^{1}$,
\newauthor{Pavel E. Mancera Piña$^{1,4,5}$ and Marc A. W. Verheijen$^{1}$}
\\
\footnotesize
$^{1}$Kapteyn Astronomical Institute, University of Groningen, Landleven 12, 9747 AD Groningen, The Netherlands\\
\footnotesize
$^{2}$INAF–Osservatorio Astronomico di Padova, Vicolo dell’Osservatorio 5, I-35122, Padova, Italy\\
\footnotesize
$^{3}$INAF–Osservatorio Astrofisico di Arcetri, Largo E. Fermi 5, I-50157, Firenze, Italy\\
\footnotesize
$^{4}$ ASTRON, Netherlands Institute for Radio Astronomy, Postbus 2, NL-7900 AA Dwingeloo, The Netherlands\\
\footnotesize
$^{5}$ Leiden Observatory, Leiden University, P.O.Box 9513, NL-2300 AA Leiden, The Netherlands\\
}

\date{Accepted January 4, 2023; Received September 14, 2022; in original form February 11, 2022}

\pubyear{2023}

\begin{document}
\label{firstpage}
\pagerange{\pageref{firstpage}--\pageref{lastpage}}
\maketitle
\begin{abstract}
 We use a dynamical model of galactic fountain to study the neutral extraplanar gas (EPG) in the nearby spiral galaxy NGC~2403. We have modelled the EPG as a combination of material ejected from the disc by stellar feedback (i.e. galactic fountain) and gas accreting from the inner circumgalactic medium (CGM). This accretion is expected to occur because of cooling/condensation of the hot CGM (corona) triggered by the fountain. Our dynamical model reproduces the distribution and kinematics of the EPG \ion{H}{i} emission in NGC~2403 remarkably well and suggests a total EPG mass of $4.7^{+1.2}_{-0.9}\times 10^8\,\mathrm{M_\odot}$, with a typical scale height of around 1\,kpc and a vertical gradient of the rotation velocity of $-10.0\pm2.7\,\mathrm{km\,s^{-1}\,kpc^{-1}}$. The best-fitting model requires a characteristic outflow velocity of $50\pm10\,\mathrm{km\,s^{-1}}$. {The outflowing gas} starts out mostly ionised and only becomes neutral later {in the trajectory}. The accretion rate from the {condensation of the} inner {hot} CGM inferred by the model is 0.8\,$\mathrm{M}_\odot\,\mathrm{yr}^{-1}$, approximately equal to the star formation rate in this galaxy (0.6\,$\mathrm{M}_\odot\,\mathrm{yr}^{-1}$). We show that the accretion profile, which peaks at a radius of about 4.5\,kpc, predicts a disc growth rate compatible with the observed value. Our results indicate that fountain-driven corona condensation is a likely mechanism to sustain star formation as well as the disc inside-out growth in local disc galaxies.
 \end{abstract}




\section{Introduction}
\label{sec:introduction}
Nearby spiral galaxies have been forming stars, across their lifetimes, at an approximately constant or gently declining rate, despite the fact that the gas in their interstellar medium (ISM) {would, without replenishment,} be consumed in a few Gyr \citep{Aumer09,Tacconi18}. An external gas reservoir is therefore needed from which galaxies accrete gas at a rate compatible with their SFR \citep[e.g.][]{Fraternali12}. Gas-rich mergers are not providing a sufficient contribution, at least in the local Universe \citep{Sancisi08, Di14}. Therefore the majority of the accretion must come from the diffuse gas that resides outside galaxies.

The multi-phase circumgalactic medium (CGM) is expected to host a significant fraction of the baryons associated with dark matter halos in normal spiral galaxies \citep[e.g.][]{Crain07,Tumlinson11,Li18}, which makes it the most probable gas reservoir eligible for accretion. {A prominent} component of the CGM is hot gas ($T\sim10^{6-7}\,$K) in the form of a diffuse `corona' at nearly the virial temperature and in {nearly} hydrostatic equilibrium with the dark matter potential \citep[e.g.][]{White91,Pezzulli17}. Galactic coronae are thought to surround {galaxies and to be extended to their virial radii} \citep{Fukugita06,Faerman20}. Direct detection of the hot coronae {in X-rays} is limited to the innermost few tens of kpc in massive {galaxies with stellar mass beyond $10^{11}\,\mathrm{M_\odot}$} \citep[e.g.][]{Anderson11, Walker15, Anderson16}, {while indirect evidence of their presence extends further \citep[e.g.][]{Gatto13,Putman21}}. Cool CGM ($T\sim10^{4}\,$K) {gas has also been detected, mostly in absorption along quasar sightlines,} in several studies \citep[e.g.][]{Heckman17,Rubin18,Zahedy19}. Like the hot corona, also these cool absorbers extend to large distances (up to and sometimes beyond the virial radius) and their origin and fate remain debated \citep{Rubin10, Schroetter19, Pointon19, Afruni21}. 

Although gas accretion from the CGM is crucial to feed star formation \citep{Hopkins08, Sancisi08, Keres09}, how precisely it takes place is still unknown. {One possible accretion scenario is that cold filaments reach the outer disc \citep{Lagos17,El-Badry18,Trapp22} and are transported into the inner star-forming regions via radial motions, although \citet{Di21} found that radial inflows in nearby galaxies alone could not sustain the star formation rates. Other possible mechanisms include cold gas filaments directly feeding the inner regions of a galaxy or the cooling of the hot corona \citep{Keres05,Nelson13,Voit15}}. The spontaneous cooling of the corona via thermal instability is still under debate as a number of works suggest that the combination of buoyancy and thermal conduction can suppress the growth of thermal perturbations \citep[e.g.][]{Binney09,Nipoti10, Joung12}. Some authors have proposed that coronal condensation could be triggered by the ejection of gas from the disc due to stellar feedback, such as in supernova-powered superbubbles \citep[][and references therein]{Fraternali17}. In this scenario, the cooling of the hot gas is due to the mixing with the cool gas ejected from the disc and occurs within the fountain cycle. This process can be detected in high-quality data as it leaves a mark in the kinematics of the ejected disc gas \citep{Fraternali08,Marasco12}.

To gain insight into the gas exchange processes between the disc and the {inner hot} CGM, one must focus on the disc-halo interface region. Deep \ion{H}{i} observations have shown that disc galaxies, including the Milky Way, are surrounded by a neutral gas layer extending up to a few kpcs from their disc planes \citep[e.g.][]{Wakker01,Sancisi08,Hess09,Marasco11}. This gas layer, known as extraplanar gas (EPG), is nearly ubiquitous in late-type galaxies and has a mass of 10--30 per cent of the mass of the \ion{H}{i} in the disc \citep{Marasco19}. The kinematics of the EPG is primarily characterised by differential rotation, similar to the disc, but with {a negative} rotational gradient {(lag)} ranging from $-10$ to $-20\,\mathrm{km\,s^{-1}\,kpc^{-1}}$ in the vertical direction \citep[e.g.][]{Oosterloo07, Zschaechner11}. Non-circular motions, especially large-scale inflows are also often found {\citep[e.g.][]{Fraternali02, Barbieri05,Marasco19}}. Ionised EPG has also been detected, both in the Milky Way \citep{Dettmar90, Lehner12,Lehner22} and in several other galaxies \citep{Heald05,Levy19}, with similar kinematics as the neutral EPG \citep{Kamphuis07,Li21,Marasco22}.
 
The similarity between EPG and disc kinematics strongly suggests that EPG originates mostly from the disc, very likely pushed out of the plane due to stellar {feedback and pulled back by gravity. This phenomenon is also known as `galactic fountain'} \citep{Shapiro76,Bregman80}. \citet[][hereafter FB06]{Fraternali06} built ballistic models of galactic fountain flows, which {successfully reproduced} many of the observed properties of the EPG in the two nearby galaxies NGC~891 and NGC~2403. {It is worth noticing that} ballistic models also describe very well the {properties} of the warm gas (neutral and ionised) in the hydrodynamical TIGRESS simulations \citep{Vijayan20}. However, a pure fountain model failed to reproduce the net inward flow {(instead, an outward flow was predicted)} and underestimated the rotation lag compared to the observed EPG { in NGC~891 and NGC~2403}. \citet[][hereafter FB08]{Fraternali08} mitigated these issues by introducing an external factor that could lower the angular momentum of fountain gas: accretion from the ambient gas. Although initially introduced to reproduce the kinematics of the EPG, the net inflow rate derived from this model turned out to be consistent with the SFR of the two galaxies, suggesting that the accretion triggered by the fountain cycle could be a viable mechanism to maintain the star formation activity. 
 
{An unsolved} issue of the above fountain-driven accretion scenario was the source of the accretion. This has been explored by \citet{Marinacci10} with hydrodynamical simulations. Their simulations of fountain gas clouds interacting with the hot corona indicated that the corona was a possible accretion source. During the interaction process, part of the fountain gas is stripped and mixed with the hot gas. The mixture has a typical temperature of $T\sim10^{5}\,$K, where the cooling function peaks, and also higher metallicity and density than the hot corona. As a consequence, the cooling time is {reduced} to a value shorter than the travel time of fountain gas. This result has been confirmed by other simulations with increasing levels of complexity \citep{Armillotta16,Gronke18,Kooij21}. {Some studies have upgraded the approach of FB08, taking into account the results of hydrodynamical simulations, using physical properties of the EPG and the hot corona as adjustable parameters, and} managed to reproduce the phase-space distribution of both neutral and ionised EPG in the disc--halo interface of Milky Way remarkably well \citep[][hereafter M12]{Marasco13, Fraternali13,Marasco12}. The best-fitting model predicted a net inflow rate which is consistent with the SFR of the Milky Way. 
 
The aforementioned studies strongly suggest that fountain-driven accretion takes place in the Milky Way and provides a promising explanation for how galaxies like our own can sustain their star formation with time. {However, so far the Milky Way remains the only galaxy for which a state-of-the-art model of the galactic fountain has been applied to the observations using a parametric fitting methodology, which is required to robustly characterise the fountain flow and to quantify the properties of the accreting gas. The earlier models in FB08 did not statistically explore the parameter space, and furthermore, did not include the condensation of the corona, since hydrodynamical simulations were not available by then.  In this paper, we {revisit this by applying our state-of-the-art fountain model} to NGC~2403, using high-quality \ion{H}{i} data {(with a beam size of 30\arcsec $\times$ 29\arcsec and an rms-noise of 0.19\,mJy$\mathrm{\,beam^{-1}}$)} from \citet{Fraternali02}, which were later included in the HALOGAS survey \citep{Heald11}.} Table~\ref{tab:info} summarises the main physical properties of NGC~2403.

{\begin{table*}
\begin{tabular}{lrrrrcccccc}
\hline
\multicolumn{1}{c}{Galaxy Name}&\multicolumn{1}{c}{RA}&\multicolumn{1}{c}{DEC}&\multicolumn{1}{c}{PA}&\multicolumn{1}{c}{INCL}&\multicolumn{1}{c}{Distance}&\multicolumn{1}{c}{Hubble Type}&\multicolumn{1}{c}{\textit{M}$_\mathrm{B}$}&\multicolumn{1}{c}{$\mathrm{M_*}$}&\multicolumn{1}{c}{$\mathrm{M_{HI,EPG}}$}&\multicolumn{1}{c}{$\mathrm{SFR}$}\\
\multicolumn{1}{c}{} & \multicolumn{1}{c}{} & \multicolumn{1}{c}{} & \multicolumn{1}{c}{[{\degr}]} & \multicolumn{1}{c}{[{\degr}]} &\multicolumn{1}{c}{[Mpc]} &
\multicolumn{1}{c}{} &\multicolumn{1}{c}{} &
\multicolumn{1}{c}{[$10^8\,\mathrm{M_\odot}$]} &
\multicolumn{1}{c}{[$10^8\,\mathrm{M_\odot}$]}&
\multicolumn{1}{c}{[$\mathrm{M_\odot}\,{\mathrm{yr^{-1}}}$]}\\
\hline
\multicolumn{1}{c}{(1)} & \multicolumn{1}{c}{(2)} & \multicolumn{1}{c}{(3)} & \multicolumn{1}{c}{(4)} & \multicolumn{1}{c}{(5)} & \multicolumn{1}{c}{(6)} & \multicolumn{1}{c}{(7)} &
\multicolumn{1}{c}{(8)} &
\multicolumn{1}{c}{(9)} &
\multicolumn{1}{c}{(10)} &
\multicolumn{1}{c}{(11)}\\
 \hline \hline
 NGC~2403& 07$^\mathrm{h}$36$^\mathrm{m}$51\fs4&+65\degr36\arcmin09\farcs2&$124.6$&$62.5$&3.2&SAcd &$-19.68$&71.9&$5.9$&$0.6$\\
\hline
\end{tabular}
\caption{Galaxy properties. Columns: (1) Galaxy name. (2)--(3): Coordinates (J2000). (4)--(5): Position-angle and inclination. (6) Distance. (7) Hubble type. (8) Absolute magnitude in the $B$-band. (9) Stellar mass \citep[see][]{Pezzulli15}. (10) Total mass of \ion{H}{i} extraplanar gas. (11) Total star formation rate of the galaxy. Values in this table are taken from \citet{Marasco19} unless otherwise mentioned.}
\label{tab:info}
\end{table*}}

In Section~\ref{sec:model} we {provide a description of our dynamical model of the galactic fountain.} In Section~\ref{sec:implementation} we discuss the customisation we have made to implement the model for the case of NGC~2403. In Section~\ref{sec:results} we present the modelling results. In Section~\ref{sec:discussion} we discuss the reliability of our results and possible {implications}. We summarise our analysis in Section~\ref{sec:conclusion}.

\section{The model}
\label{sec:model}

In this Section, we describe the main ingredients of our model and discuss its main free parameters. Further details can be found in FB06, FB08 and M12. We consider two different types of models: a `pure fountain' ballistic model and a `fountain + corona accretion' model which takes { the interaction of fountain clouds} with the hot coronal gas into consideration. {In both scenarios, the models have a quasi-stationary state and are axisymmetric}. The neutral EPG in the disc--halo interface region is modelled as a collection of clouds that are ejected from the disc at different radii with a given distribution of initial velocities { and angles}, and whose orbits are then integrated in time and followed across the halo region until they return to the disc.

{Since galactic fountains are powered by stellar feedback, we assume that the amount of gas ejected from each location in the disc is proportional to the SFR surface density at that radius. In practice, we incorporate this assumption by assigning, to each of our modelled clouds, {a weight} proportional to the SFR surface density at the ejection radius. This {weight} is then {factored in when creating} the mock datacube to be compared with observations (see also further explanations below).}

 {In our pure fountain ballistic models}, the trajectories of the fountain clouds are integrated using a numerical approximation of the galaxy gravitational potential, derived as described in Section~\ref{sec:gravity}. For fountain + corona accretion models, hydrodynamical forces due to the interaction between the clouds and the hot corona are parameterised in simple forms described in Section~\ref{sec:model_accretion}. 

The positions and velocities of the clouds along their orbits are recorded at each time-step (0.3\,Myr), projected along the line-of-sight of the observer, {weighted by the local SFR surface density at the ejection radius} and transferred into a synthetic datacube, which is then adapted to a specific galaxy (NGC~2403 in our case) by assuming a distance, inclination (INCL), and position angle (PA), and using the same observational setup (beam shape, spectral resolution, pixel size, etc.) of the data under consideration. The outcome of the dynamical model is therefore a synthetic datacube which can be directly compared with the observational \ion{H}{i} data of our target galaxy.

Construction of the model involves several parameters but we will focus preferentially on three (for pure fountain models) or four (only for fountain + corona accretion models) that regulate the initial outflow speed of the clouds, their neutral gas fraction, the EPG total mass and, for models that include interaction with the corona, an additional parameter that regulates the condensation efficiency. Below we discuss these parameters in detail. Other ingredients are fixed by the observations, in particular the galaxy potential (which affects the trajectory of the cloud) and the SFR surface density profile (which regulates the ejection rate), as described in Section~\ref{sec:implementation}.

\subsection{Outflow velocity}
Fountain clouds are initially located within the galaxy disc and rotate at the circular speed set by our gravitational potential\footnote{They also feature an additional velocity component, with an amplitude randomly extracted from a Gaussian distribution with rms of 8$\,\mathrm{km\,s^{-1}}$ and a random (isotropic) direction, to simulate the typical velocity dispersion of the neutral ISM \citep{Iorio17,Bacchini19,Mancera21}.}. Each cloud receives a `kick' with a velocity $v_\mathrm{k}$ at certain angles $\theta$, {which is defined as the angle between the velocity vector and the direction normal to the disc plane.} The probability distribution of the ejection as a function of $v_\mathrm{k}$ and $\theta$ {(assuming a uniform probability in the azimuthal direction)} follows FB06 and is given by
\begin{equation}
\mathcal{P}(v_\mathrm{k},\theta) \propto 
  \exp\left({-\frac{v_k^2}{2h_{v}^{2}\cos^{2\Gamma}{\theta}}}\right),
\label{eq:vel}
\end{equation}
where $h_v$ is the characteristic velocity, and $\Gamma$ determines the {level of collimation of the ejected clouds}. Larger values of $h_v$ increase the probability that a cloud is kicked at high speed. The larger $\Gamma$, the more collimated the ejection. FB06 have tested models with different values for $\Gamma$ and found that more collimated ejections agree better with the data. We have therefore fixed $\Gamma=10$ (highly collimated).

The outflow velocity of a cloud {affects} the maximum height and the trajectory of the orbit and therefore influences the final model. We, therefore, let the characteristic velocity $h_v$ be a free parameter { with a flat prior} in the range 40--$100\,\mathrm{km\,s^{-1}}$.
This range covers the typical characteristic ejection speeds of {the warm gas in} {high-resolution hydrodynamical simulations of} galactic fountains \citep{Kim18}. It also agrees with theoretical estimates of the typical blow-out speed of individual superbubbles \citep[e.g.][]{MacLow88, Keller14}.

\subsection{Phase change} 

Previous studies have found that the neutral EPG in some spiral galaxies (including the Milky Way) shows {a tentative preference for vertical inflow} \citep[for example]{Marasco19,French21}, which can be interpreted as due to a change of phase during the fountain cloud orbit: gas is largely ionised when ejected from the star-forming region of the disc but later recombines and becomes visible in \ion{H}{i} at some point during its trajectory. To account for this effect in our model, we assume that a cloud is only visible in the \ion{H}{i} phase when
\begin{equation}
v_z(t) < v_{z,0} (1- f_{\mathrm{ion}}),
\label{eq:fion}
\end{equation}
where $v_z$ is the vertical velocity (that is, in the direction perpendicular to the disc) of the cloud, $v_{z,0}$ is the vertical component of the initial outflow velocity and $f_\mathrm{ion}$ is the ionisation fraction parameter, which { we set as a free parameter with a flat prior and} varies from zero to one. When $f_\mathrm{ion}$ equals zero, the cloud is visible in the whole orbit, while when $f_\mathrm{ion}$ equals one, the cloud is only visible when $v_z <0$ (i.e., the descending stage).


\subsection{Interaction with the corona}
\label{sec:model_accretion}

 In our model, the hot corona is modelled as a smooth, volume-filling gas layer that rotates at a lower speed than the disc, which is justified on both observational \citep{Hodges-Kluck16} and theoretical \citep{Pezzulli17} grounds. {We assume that the corona maintains a temperature of $\sim 10^6$\,K, which implicitly implies some heating by either supernova feedback \citep[e.g.][]{Stinson13} or active galactic nucleus feedback \citep[for galaxies with ongoing AGN activities; e.g.][]{Ciotti12}. The condensation and accretion of the hot corona is triggered by the cool ($T\sim10^4\,\mathrm{K}$) fountain clouds ejected from the disc, which mix efficiently with the former and produce a mixture at $T\sim10^5\,\mathrm{K}$, dramatically reducing the cooling time of the hot corona. The above processes have been investigated in the hydrodynamical simulations of cloud--corona interactions \citep{Marinacci10}. A follow-up analysis \citep{Marinacci11}} indicate that there is a net transfer of momentum from the fountain to the corona until the relative velocity between these two, $v_\mathrm{rel}$, reaches a certain threshold $v_\mathrm{thres}$. {\citet{Marinacci11} suggested $v_\mathrm{thres}\approx 75\,\mathrm{km\,s^{-1}}$ for initial conditions valid for the Milky Way but pointed out that $v_\mathrm{thres}$ can vary in the range 45--105\,$\mathrm{km\,s^{-1}}$ \citep[see also][]{Fraternali17}}. {As soon as $v_\mathrm{rel}$ becomes smaller than this threshold $v_\mathrm{thres}$,} the net momentum transfer ceases as the condensation of corona recaptures angular momentum lost by fountain gas. For this reason, we set the azimuthal speed of the corona to be always lower than the local circular speed $v_\mathrm{c}$ by $v_\mathrm{thres}$, and in this case, $v_\mathrm{c}-75\,\mathrm{km\,s^{-1}}$. In Section~\ref{sec:lagexpri} we explore models with different value of $v_\mathrm{thres}$, corresponding to different rotational speeds for the coronal gas.
 
 In the above scenario, the cloud acceleration due to interaction with the corona is defined as
\begin{equation}
\boldsymbol{\dot{v}} =
\left\{
\begin{array}{lr}
 -\frac{\mathrm{C}\rho_\mathrm{hot}\sigma_\mathrm{cloud}(v_\mathrm{rel}-v_\mathrm{thres})}{M_\mathrm{cloud}}\,\boldsymbol{v_\mathrm{rel}} - \alpha\boldsymbol{v_\mathrm{rel}},&{v_\mathrm{rel}\,\ge\,v_\mathrm{thres}}\\ 
  - \alpha\boldsymbol{v_\mathrm{rel}},&{v_\mathrm{rel}\,<\,v_\mathrm{thres}},
  \end{array}
\right.
\label{eq:conden}
\end{equation}
where $\boldsymbol{v_\mathrm{rel}}$ is the cloud-corona relative velocity vector, $v_\mathrm{rel}$ is the modulus of $\boldsymbol{v_\mathrm{rel}}$, $M_\mathrm{cloud}$ and $\sigma_\mathrm{cloud}$ are the mass and the cross-section of the cloud (defined as $\pi{R_\mathrm{cloud}^2}$, {with $R_\mathrm{cloud}$ the radius of the cloud}), $\rho_\mathrm{hot}$ is the density of the corona, C is a dimensionless constant of order unity { (in our model C=1)} to account for the geometry of the cloud, and $\alpha$ is the condensation rate of the coronal gas onto the cloud, such that the mass of the cloud $M_\mathrm{cloud}$ grows with time as $\dot{M}_\mathrm{cloud}={\alpha}M_\mathrm{cloud}$. We assume a corona density of $10^{-3}\,\mathrm{cm}^{-3}$, a cloud radius of 100\,pc and an initial mass of 
2$\times\,10^4\,\mathrm{M_\odot}$, consistent with typical values of fountain clouds suggested by observations \citep{Hsu11}. 

The first term on the right-hand side of equation~\ref{eq:conden} represents the drag experienced by the fountain cloud as it moves through the coronal gas: the cloud speed decreases as long as its velocity stays above $v_\mathrm{thres}$. The second term is due to the condensation of coronal gas onto the cloud: as the total mass of the cloud increases, conservation of the total momentum implies lower velocity \citep[see][]{Fraternali08}. We have also derived the drag timescale  $t_\mathrm{drag}= 724$\,Myr using equation(6) in \citet{Fraternali17}, which is larger than the fountain orbit time ($\sim$100\,Myr), we therefore expect that drag only has a minor effect.

In fountain + corona accretion models, we let $\alpha$ be a free parameter {with a flat prior} in the range $\alpha=0$--6 Gyr$^{-1}$.

\subsection{EPG mass}

The normalisation of the {\ion{H}{i} flux presented in the} final galactic fountain model sets the total \ion{H}{i} EPG mass, which is another free parameter. We use a fiducial EPG mass of $5.9\times10^8\,\mathrm{M_\odot}$ from \cite{Marasco19} as an initial guess, but allow the EPG mass to vary, multiplying the fiducial value by a normalisation scaling factor in the range 0.1--10.

\section{Implementation of the model}
\label{sec:implementation}

In this section, we describe the gravitational potential and the SFR surface density radial profile for NGC~2403, as they are necessary ingredients to construct our dynamical models. We then describe how we fit the model parameters to the data.
\subsection{The gravitational potential}
\label{sec:gravity}
{We use the gravitational potential grid derived by FB06 for NGC~2403 without modification. Below we briefly describe how the potential model is built. 

The gravitational potential was derived from an axisymmetric mass model, which consists of three components: a stellar disc, a gaseous disc, and an NFW dark matter halo \citep{Navarro97}. FB06 performed a mass decomposition of the \ion{H}{i} rotation curve of NGC~2403 \citep{Fraternali02} using the three components mentioned above. The stellar and the gaseous discs' density distributions were given by exponential profiles, along both the radial ($R$) and the vertical ($z$) direction. {The scale length of the stellar (gaseous) disc $R_\mathrm{*}$ ($R_\mathrm{gas}$) was derived by fitting an exponential profile to the stellar (gaseous) surface brightness radial profile. The scale height of the stellar disc was set to one-fifth of its scale length (see \citealt{Kruit11} and references therein), and the scale height of the gaseous disc was set to 100\,pc \citep[typical of the inner gaseous disc, see][]{Marasco17b,Bacchini19,Mancera22}. The mass-to-light ratio of the stellar disc was derived via the rotation curve decomposition}. The above parameters of the mass model are listed in Table~\ref{tab:sffr}. Once the parameters of all components are decided, the galactic potential and forces are calculated numerically in the $(R, z)$ cylindrical coordinate system, using a grid with a cell size of 0.1\,kpc within $R < 25$\,kpc and $z < 5$\,kpc,  and of 0.5\,kpc for $25 < R < 100$\,kpc and $5 < z < 100$\,kpc. {Potential and forces are determined at any ($R$,$z$) via a bilinear interpolation of these grids (see FB06 for details).}

\begin{table}
\begin{tabular}{ccccccccc}
\hline
\multicolumn{1}{c}{$(M/L)_\mathrm{*}$}&\multicolumn{1}{c}{$R_\mathrm{*}$}&\multicolumn{1}{c}{$h_\mathrm{*}$}&\multicolumn{1}{c}{$R_\mathrm{gas}$}&\multicolumn{1}{c}{$h_\mathrm{gas}$}&\multicolumn{1}{c}{${\rho}_{\mathrm{0,DM}}$} &\multicolumn{1}{c}{r$_s$}\\
\multicolumn{1}{c}{} & \multicolumn{1}{c}{[kpc]} & \multicolumn{1}{c}{[kpc]} & \multicolumn{1}{c}{[kpc]} & \multicolumn{1}{c}{[kpc]} &
\multicolumn{1}{c}{[M$_\odot\,\mathrm{kpc^{-3}}$]} &
\multicolumn{1}{c}{[kpc]}\\
\hline
\multicolumn{1}{c}{(1)} & \multicolumn{1}{c}{(2)} & \multicolumn{1}{c}{(3)} & \multicolumn{1}{c}{(4)} & \multicolumn{1}{c}{(5)} & \multicolumn{1}{c}{(6)} &
\multicolumn{1}{c}{(7)} \\
 \hline \hline
1.70&2.0&0.4&5.7&0.1&$3.1\times10^7$&$4.5$\\
\hline
\end{tabular}
\caption{Mass models for NGC~2403. Columns: (1) Mass-to-light  ratio {in the $B$-band} of the stellar disc. (2)--(3): Scale length and scale height of the stellar disc. (4)--(5): Scale length and scale height of the gaseous disc. (6)--(7) Central density and scale radius of the NFW dark matter halo.}
\label{tab:sffr}
\end{table}}

\subsection{Star-formation-rate surface-density profiles}
\label{sec:sfr_profile}

 In this paper, we directly use the SFR surface density radial profiles from previous observations, as opposed to FB06, which used the Schmidt–Kennicutt law \citep{Kennicutt89}, and M12, {which used another empirical star formation law (directly derived from 17} galaxies with known gas and SFR surface densities) to estimate the SFR.

\begin{figure}
\centering
\includegraphics[scale=0.51]{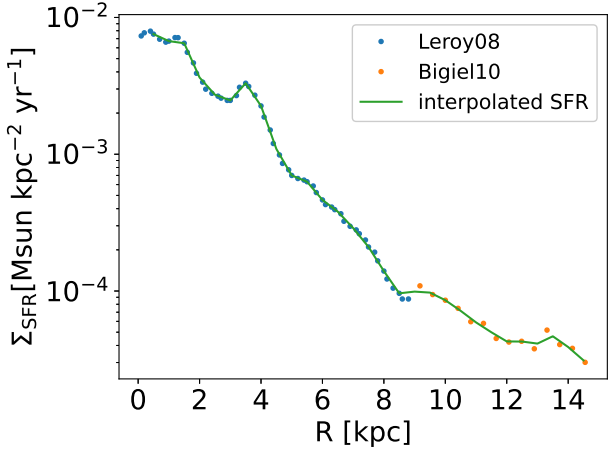}
\caption{Star formation rate surface density versus galactocentric distance in NGC~2403. Blue dots represent data from \citet{Leroy08} while orange points are from \citet{Bigiel10}. The green curve shows the interpolated profiles with steps of 0.5\,kpc and is used as an input for our fountain model.}
\label{fig:sfr}
\end{figure}

The SFR surface-density profile of NGC~2403 is mainly taken from \cite{Leroy08}, which derived the SFR using a combination of far ultraviolet (FUV) and 24\,$\mu$m data, {and is then complemented with the SFR surface density profile from \cite{Bigiel10}, which is derived from FUV data with a lower resolution but larger radial extent compared to \citet{Leroy08}}. We refer the readers to \citet{Bacchini19,Bacchini20} for more details about collecting SFR data of NGC~2403. Fig.~\ref{fig:sfr} shows the SFR surface-density data and the interpolated profile (in steps of 0.5\,kpc) which we used as an input for our fountain models.

\subsection{Separation of the EPG emission}
\label{sec:mask}

{Before modelling the EPG in the NGC~2403 datacube, we first need to isolate its emission from the underlying disc and from external regions {(foreground and background emission)} that are clearly not associated with the galaxy. For this purpose, we follow the procedure described in \cite{Marasco19}. 

The emission from regions external to the galaxy is filtered out by spatially smoothing the datacube by a 2D Gaussian kernel with a full width half maximum (FWHM) of $64\farcs5\times54\farcs6$, which is five times larger than the {spatial} resolution of the data}, calculating a smoothed rms noise level, and then sigma-clipping at $\mathrm{S/N}=4$. This produces a mask that is applied to the original (not smoothed) data to exclude the regions external to the main galaxy.

In intermediate-inclination galaxies like NGC~2403, the emission from the EPG overlaps spatially with that from the regularly rotating disc but can be (at least in part) separated from the latter in the velocity space, provided that the velocity resolution is sufficient. {Here, we employ the disc--EPG separation method introduced by \citet{Fraternali02}, which works as follows. For any given \ion{H}{i} velocity profile at a certain location in the sky}, {the disc component is assumed to be described by a Gaussian profile. The EPG adds a wing to the profile, which is typically due to the lagging of EPG and located toward the systemic-velocity side; although} wings on both sides can be seen at some spatial locations across the disc due to other non-circular (mostly vertical) motions \citep[see also][]{Boomsma08}.
Despite the disc and EPG profiles are blended together, it is reasonable to neglect the contribution of the EPG around the peak of each velocity profile since EPG mass is only a small percentage ($\sim20$ per cent for NGC~2403, \citealt{Marasco19}) of the total \ion{H}{i} mass. We therefore use the `peak' region to fit the disc emission by performing a Gaussian fit using only the upper 40 per cent (in intensity) of the line profile. This Gaussian profile is considered to be the contribution of emission from the disc component alone. Pixels with disc emission {(estimated from the Gaussian profile)} larger than $N$ times the rms noise are clipped {(see \citealt{Marasco19} and \citealt{Li21} for a more detailed explanation of this methodology)}. The scaling factor $N$ is decided empirically as a compromise between keeping enough EPG emission for the modelling and alleviating the disc contamination. We set $N=2$ for NGC~2403. 

Some peculiar features in NGC~2403, in particular, a long filament of unknown origin \citep[see also][]{deblok14} have also been manually filtered out (see blank regions in Figs.~\ref{fig:pvn2403gf} and \ref{fig:pvn2403ac}). We discuss this further in Section~\ref{sec:lagexpri}.

After passing through the above mask, only EPG emission and noise remain in the datacube. We then implement sigma-clipping at $\mathrm{S/N}=2$ to mask the random noise. For consistency, the same mask has also been applied to the model datacube that we describe below.

\subsection{Model construction and evaluation}
\label{sec:model_eval}

Our EPG models have three or four free parameters: the characteristic outflow velocity $h_v$, the ionisation fraction $f_\mathrm{ion}$, the condensation rate $\alpha$ (for fountain + corona accretion models), and the EPG mass M$_\mathrm{EPG}$. We build three(four)-dimensional grids for pure fountain (fountain + corona accretion) models with $h_v$ varying from 40 to $100\,\mathrm{km\,s^{-1}}$ in steps of $10\,\mathrm{km\,s^{-1}}$, $f_\mathrm{ion}$ varying from 0.0 to 1.0 in steps of 0.2, $\alpha$ varying from 0 to 6\,Gyr$^{-1}$ in steps of 0.6\,Gyr$^{-1}$, and scaling factor of the initial EPG mass varying from 0.1 to 10 in steps of factor of $10^{0.2}$. {The ranges and steps of the free parameters are summarised in Table~\ref{tab:free_param}.}

\begin{table*}
\begin{tabular}{llccl}
\hline
\multicolumn{1}{c}{Parameter}&
\multicolumn{1}{c}{description}&
\multicolumn{1}{c}{range}&
\multicolumn{1}{c}{step}&
\multicolumn{1}{c}{units}\\
\hline
 $h_v$&Characteristic outflow velocity (equation~\ref{eq:vel})&[40,100]&10&$\mathrm{km\,s^{-1}}$\\
 $f_\mathrm{ion}$&Ionisation fraction during the ascending part of the orbits(equation~\ref{eq:fion})&[0,1.0]&0.2& \\
 $\alpha$&condensation rate of coronal gas (equation~\ref{eq:conden})&[0,6.0]&0.6&Gyr$^{-1}$\\
 Norm&EPG mass scaling factor $^{a}$&[0.1,10]&$10^{0.2}$& \\
\hline
\end{tabular}
\caption{Free parameters of our galactic fountain model. The third column lists the range explored in our residual calculations, using a grid size given by the forth column. $^{a}$ a value of $1$ corresponds to the EPG mass determined by \citet{Marasco19} ($5.9\times 10^8\,\mathrm{M_\odot}$).}
\label{tab:free_param}
\end{table*}

The best-fitting parameters are estimated by a Bayesian approach. 
For each cell in our 3D (4D) parameter grid, we compute the posterior probability of our model. For a chosen parameter vector $\mathbf{x}$ and given our data $\mathcal{D}$, the posterior probability $\mathcal{P}$ is given by 
\begin{equation}
 \mathcal{P}(\mathbf{x}\vert\mathcal{D})\propto\mathcal{P}(\mathcal{D}|\mathbf{x})\mathcal{P}(\mathbf{x}),
\end{equation}
\noindent where $\mathcal{P}(\mathcal{D}|\mathbf{x})$ is the likelihood function and $\mathcal{P}(\mathbf{x})$ is the prior. The prior for each parameter is uniform within the parameter space {(uniform in the logarithmic scale for the normalisation parameter)}. The likelihood function is given by 
\begin{eqnarray}
\mathcal{P}(\mathcal{D}\vert\mathbf{x}) &\propto& \prod\limits_{n.voxels}^{} \exp\left({-\frac{|\mathcal{M}(\mathbf{x})-\mathcal{D}|}{\varepsilon}}\right) \nonumber \\ 
 &=&\exp\left({-\sum\limits_{n.voxels}^{}\frac{|\mathcal{M}(\mathbf{x})-\mathcal{D}|}{\varepsilon}}\right)\nonumber\\ 
 &=& \exp[-\mathcal{R}(\mathbf{x})/\varepsilon],
\label{eq:likelihood}
\end{eqnarray}
\noindent where $\mathcal{M}$ represents the model datacube built from parameter vector $\mathbf{x}$, $\varepsilon$ is the uncertainty of the data, and $\mathcal{R}$ is the sum of the absolute residuals between the data and the model, which is defined as the sum of absolute difference in each pixel: $\mathrm{Res}=\sum|\mathrm{data}-\mathrm{model}|$. Note that both the model and the data have been masked using the method described in Section~\ref{sec:mask}, i.e, {the voxels where EPG emission is detected at more than $2\sigma$ are considered in the determination of the residuals}. {In equation~\ref{eq:likelihood}, $\varepsilon$ regulates how rapidly the likelihood drops when our model deviates from the data.
Assuming $\varepsilon$ equal to the rms-noise of the data is a poor choice, which leads to very narrow posterior probability distributions and severely underestimates the uncertainties in our model parameters. This occurs because our model is smooth and axisymmetric, and cannot possibly capture the complexity of the data down to the noise level. Numerical solutions to this problem can be worked out \citep[see Section 2.5 in][]{Marasco19}, but in this work, we prefer to set $\varepsilon$ a posteriori, in a way that the 2-$\sigma$ uncertainty on the derived parameters corresponds to models that look very different from the data by visual inspection.} {In the end,} we assume $\varepsilon\,=\,0.38\,\mathrm{Jy\,beam^{-1}}$. We marginalise the {multi-dimensional posterior distribution} to determine the probability distribution of individual parameters. Best-fitting values are defined as the median of these marginalised posterior distributions, and the uncertainties are taken as half the difference between the 84th and 16th percentiles of the distribution.

\section{Results}
\label{sec:results}

\subsection{Residuals and position-velocity diagrams}
\label{sec:results_n2403}

In this Section, we show the best-fitting results of the pure fountain and the fountain + corona accretion models. {The 2D marginalised posterior probability distributions} are shown in Appendix~\ref{appen:a}. The best-fitting values and uncertainties, obtained with the method described in Section~\ref{sec:model_eval}, are listed in Table~\ref{tab:bestfit}. .

\begin{table*}
\begin{tabular}{lrrrrrrrr}
\hline
\multicolumn{1}{c}{Model}&
\multicolumn{1}{c}{$v_\mathrm{thres}$}&
\multicolumn{1}{c}{$h_v$}&
\multicolumn{1}{c}{$f_\mathrm{ion}$}&
\multicolumn{1}{c}{$\alpha$}&
\multicolumn{1}{c}{$\dot m$}&
\multicolumn{1}{c}{M$_\mathrm{EPG}$}&
\multicolumn{1}{c}{$-\ln{\mathcal{L}}$}&
\multicolumn{1}{c}{BIC}\\
\multicolumn{1}{c}{} & \multicolumn{1}{c}{[$\mathrm{km\,s^{-1}}$]} &\multicolumn{1}{c}{[$\mathrm{km\,s^{-1}}$]} &
\multicolumn{1}{c}{} & \multicolumn{1}{c}{[Gyr$^{-1}$]} & 
\multicolumn{1}{c}{[$\mathrm{M}_\odot\,\mathrm{yr}^{-1}$]} & 
\multicolumn{1}{c}{[$10^8\,\mathrm{M_\odot}$]}&
\multicolumn{1}{c}{}&\multicolumn{1}{c}{}
\\
\hline
\multicolumn{1}{c}{(1)} & \multicolumn{1}{c}{(2)} & \multicolumn{1}{c}{(3)} & \multicolumn{1}{c}{(4)} & \multicolumn{1}{c}{(5)} & \multicolumn{1}{c}{(6)}& \multicolumn{1}{c}{(7)} & \multicolumn{1}{c}{(8)}&
\multicolumn{1}{c}{(9)}\\
 \hline \hline
 pure fountain&$N/A$&$50\pm10$&0.6$\pm0.2$&$N/A$&$N/A$&$5.9^{+1.5}_{-1.2}$&232.6&490.6\\
 \\
 fountain + corona accretion&75&$50\pm10$&0.4$\pm0.2$&2.4$^{+1.8}_{-0.6}$&0.8$^{+0.4}_{-0.2}$&$4.7^{+1.2}_{-0.9}$&224.5&482.9\\ \\
 fountain + corona accretion&45&$50\pm10$&0.4$^{+0.2}_{-0.4}$&4.2$\pm1.2$&1.1$^{+0.3}_{-0.2}$&$4.7^{+1.2}_{-0.9}$&223.5&480.9\\
\hline
\end{tabular}
\caption{The best-fitting values and uncertainties (obtained with the method described in Section~\ref{sec:model_eval}) for our fountain (+ corona accretion) models of the EPG of NGC~2403. We focus on the first two models in this Section and further discuss the third model in Section~\ref{sec:lagexpri}. (1) Model type. (2) {The velocity threshold for fountain + corona accretion models. The net transfer of momentum from the fountain to the corona ceases when the relative velocity between these two decreases below this threshold (see Section~\ref{sec:model_accretion}).} (3) Characteristic {outflow} velocity. (4) Ionisation fraction of the fountain gas. (5) Condensation rate of the hot gas. (6) Global accretion rate of the condensed hot gas onto the disc. {Note that this is not a free parameter but a value derived from the best-fitting model}. (7) {\ion{H}{i}} EPG mass. (8) Logarithm of the likelihood values $\mathcal{P}(\mathcal{D}|\mathbf{x})$ of the best-fitting models, calculated in equation~\ref{eq:likelihood}. {(9) The BIC values of  the best-fitting models, calculated from equation~\ref{eq:bic}.}}
\label{tab:bestfit}
\end{table*}

\begin{figure*}
\centering
\includegraphics[scale=0.4]{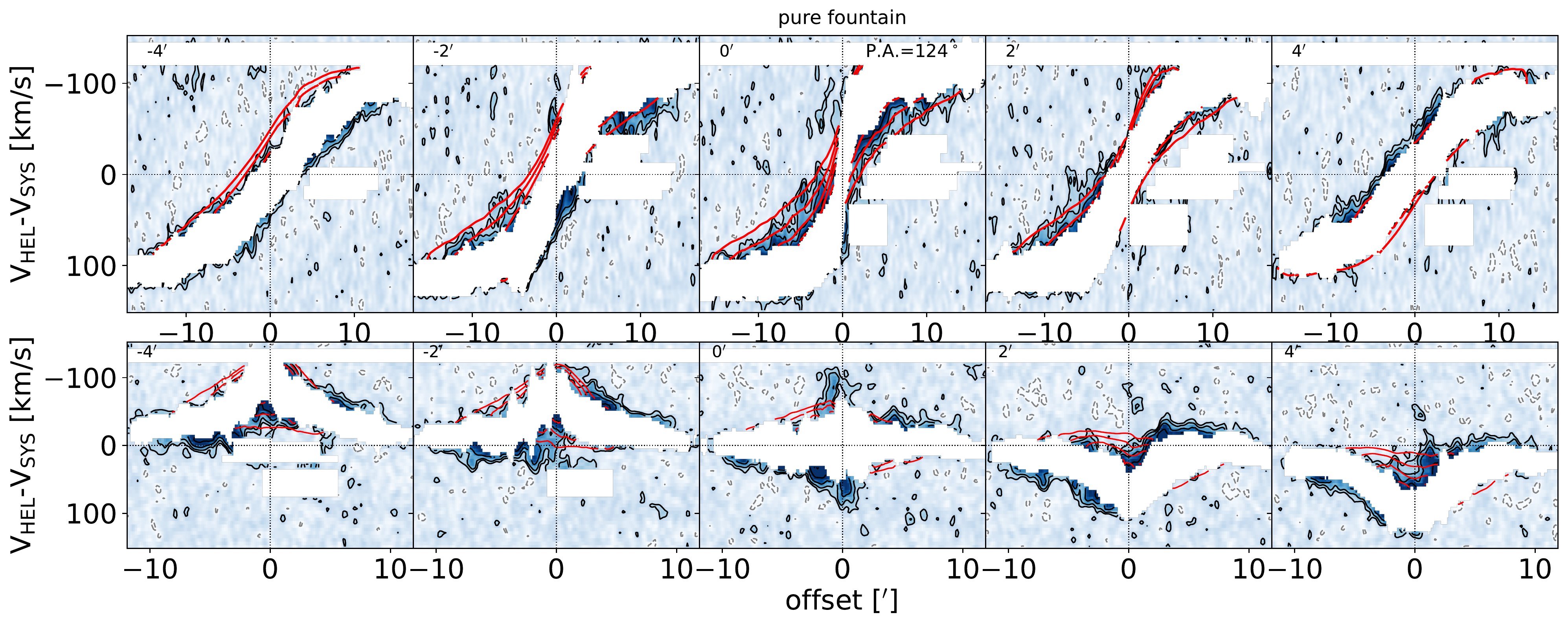}
\caption{Position--velocity (pv) slices from the data {(shown in black contours and blue colour scale)} and from the best-fitting pure fountain model (red contours); from outer to inner regions, contour levels are (2, 4, 8, 16)-$\sigma$, respectively, {and a negative contour -2$\sigma$ is shown as the dashed grey contour.}. The (irregular) blank region represents the disc mask and the square blank region represents the manual mask that filters out the irregular filament in NGC~2403. Top panels are pv slices parallel to the major axis with offsets $-4\arcmin$, $-2\arcmin$, $0\arcmin$, $2\arcmin$, $4\arcmin$. Bottom panels are pv slices parallel to the minor axis with offsets $-4\arcmin$, $-2\arcmin$, $0\arcmin$, $2\arcmin$, $4\arcmin$.}
\label{fig:pvn2403gf}
\end{figure*}

\begin{figure*}
\centering
\includegraphics[scale=0.4]{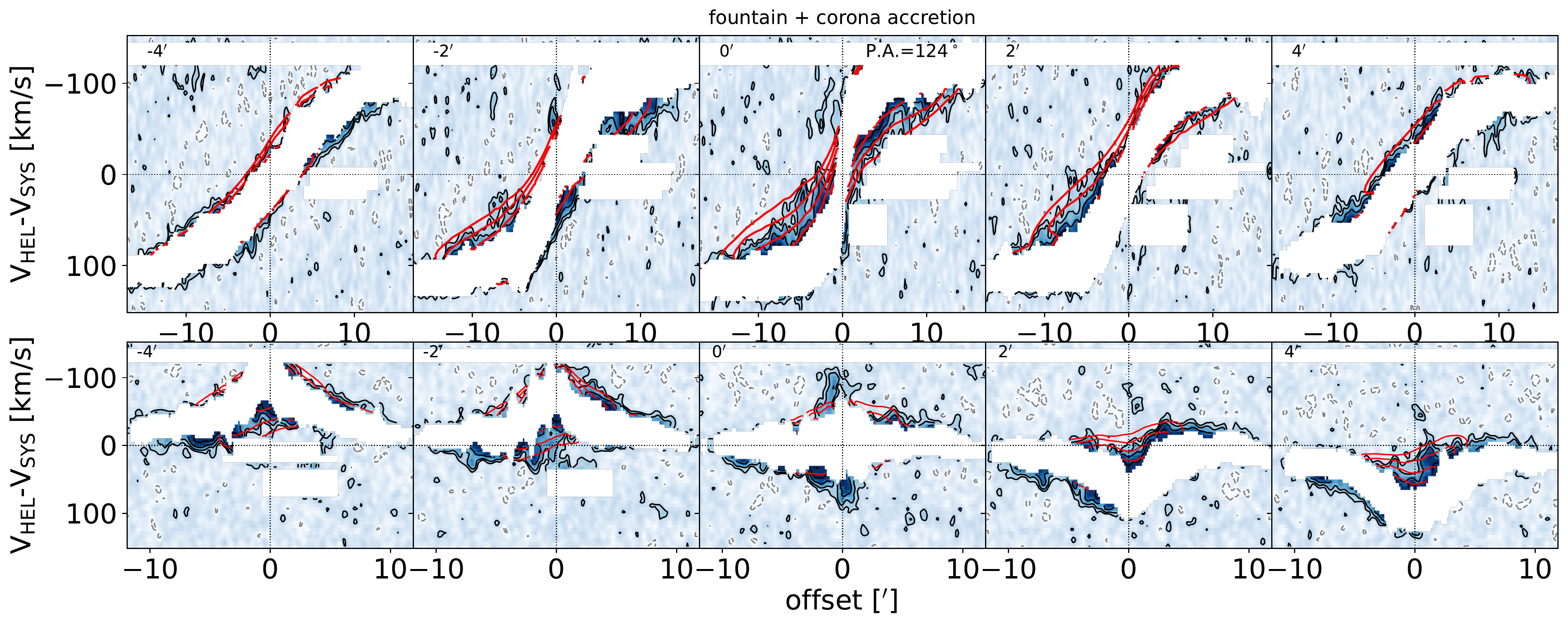}
\caption{As in Fig.~\ref{fig:pvn2403gf}, but for the best-fitting fountain + corona accretion model of NGC~2403. }
\label{fig:pvn2403ac}
\end{figure*}

\begin{figure*}
\centering
\includegraphics[scale=0.4]{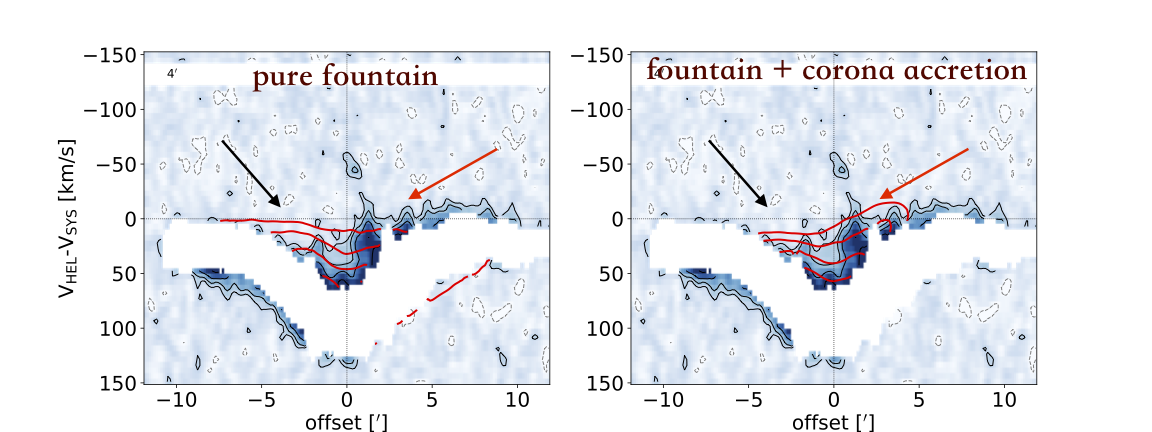}
\caption{As in Figs.~\ref{fig:pvn2403gf} and \ref{fig:pvn2403ac}, but focusing on the pv slice parallel to the minor axis with offset $4\arcmin$. Left: best-fitting pure fountain model. Right: best-fitting fountain + corona accretion model. The red arrows {mark regions where EPG emission is present in the data and in the fountain + corona accretion model, but not in the pure fountain model.} The black arrows mark out the region where the pure fountain model predicts extra emission {with respect to the data, while the fountain + corona accretion model correctly predicts a lack of emission}.}
\label{fig:pvn2403compare}
\end{figure*}

The position--velocity (pv) slices of the best-fitting models are compared with the data in Figs.~\ref{fig:pvn2403gf} and \ref{fig:pvn2403ac}. In general, both the pure fountain and fountain + corona accretion models recover the EPG emission, but we find that the former reproduces the data poorly for pv slices parallel to the minor axis. Instead, the fountain + corona accretion model performs better in the same locations. This is better shown in Fig.~\ref{fig:pvn2403compare} where we compare  the two models for a pv slice parallel to the minor axis with an offset $4\arcmin$ from the centre. The best-fitting pure fountain model fails to reproduce the emission marked out by the red arrow and predicts extra emission in the blank region marked out by the black arrow. Instead, the best-fitting fountain + corona accretion model generates the same asymmetry shown by the data. {Previous studies \citep{Fraternali02,Marasco19} have shown that this asymmetric feature can be produced by} radial inflows. In a fountain model, EPG emission shows outward radial flows, but accretion from low-angular momentum material can invert this trend and produce an inward flow (especially evident for clouds ejected from the outer regions of the disc; \citealt{Fraternali17}), which is required to best reproduce the data.

The above visual comparison prefers the fountain + corona accretion model. {This result has been already inferred by FB08, but we now have its statistical confirmation using the likelihood values} derived by equation~\ref{eq:likelihood}. We find $-\ln{[\mathcal{P}(\mathcal{D}|\mathbf{x})]}=232.6$ for the best-fitting pure fountain model, while $-\ln{[\mathcal{P}(\mathcal{D}|\mathbf{x})]}=224.5$ for the best-fitting fountain + corona accretion model, as shown in Table~\ref{tab:bestfit}. We use the Bayesian information criterion \citep[BIC;][]{Schwarz78} {to infer which of the two different scenarios (pure fountain or fountain + corona accretion) is statistically preferred by the data, given that they make use of a different number of free parameters.} The BIC is derived as 
\begin{eqnarray}
\mathbf{BIC} = -2 \ln{\mathcal{L}} + k\ln{\mathcal{N}},
\label{eq:bic}
\end{eqnarray}
where $\mathcal{L}$ is the likelihood of the model (equation~\ref{eq:likelihood}), $k$ is the number of parameters estimated by the model, and $\mathcal{N}$ is the number of independent data points used in the fit. {When comparing similar models with different numbers of free parameters, a model with a lower BIC is to be preferred, as the BIC penalises extra parameters that do not significantly lower the likelihood.} The $\mathbf{BIC}$ for the pure fountain model is 490.6 while for the accretion model is 482.9, indicating that the fountain + corona accretion model is statistically preferred by BIC.

The above results show that the \ion{H}{i} EPG of NGC~2403 is constituted by a combination of material ejected from the disc by stellar feedback and gas cooling from the inner hot CGM and accreting onto the disc. This is also consistent with previous indication from kinematic modelling of the EPG which shows radial and vertical inflow \citep{Marasco19}. {The best-fitting fountain + corona accretion model requires an outflow with a characteristic velocity of $50\pm10\,\mathrm{km\,s^{-1}}$, starting out mostly ionised and becoming neutral when the vertical velocity has been reduced by around 40\%. The inferred \ion{H}{i} total mass of the EPG ($4.7^{+1.2}_{-0.9}\times 10^8\,\mathrm{M_\odot}$) is similar to that derived in \citet{Marasco19} ($5.9\times 10^8\,\mathrm{M_\odot}$).} The accretion rate given by our best-fitting model ($0.8^{+0.4}_{-0.2}$\,$\mathrm{M}_\odot\,\mathrm{yr}^{-1}$) is compatible with the star formation rate of NGC~2403  \citep[0.6\,$\mathrm{M}_\odot\,\mathrm{yr}^{-1}$; ][]{Heald12}{\footnote{This estimate has an uncertainty of around $\pm$0.3\,dex or better, based on the algorithm \citet{Heald12} used to derive the SFR \citep{Kennicutt09}.}}, indicating that the mechanism of fountain-driven gas accretion can sustain the ongoing star formation in NGC~2403. {It is noteworthy that the values of both outflow speed and accretion rate found with our statistical analysis are in agreement with those found by FB08 by trial and error. The present analysis, however, allows us to further our understanding of fountain-driven accretion in NGC~2403.}

\subsection{Properties of the extraplanar gas layer in NGC~2403}

This is the first time that a dynamical fountain model including corona condensation has been applied to an external galaxy with a statistical fitting method. The best-fitting fountain + corona accretion model reproduces most of the EPG features in NGC~2403. Assuming our model is reliable and correct (see discussion in Section~\ref{sec:lagexpri}), we can therefore extract physical properties of the EPG layer, {as well as a predicted gas accretion profile, from the model.} 

\subsubsection{Thickness of the neutral extraplanar gas layer}
\label{sec:scaleheight}
\begin{figure}
\centering
\includegraphics[scale=0.6]{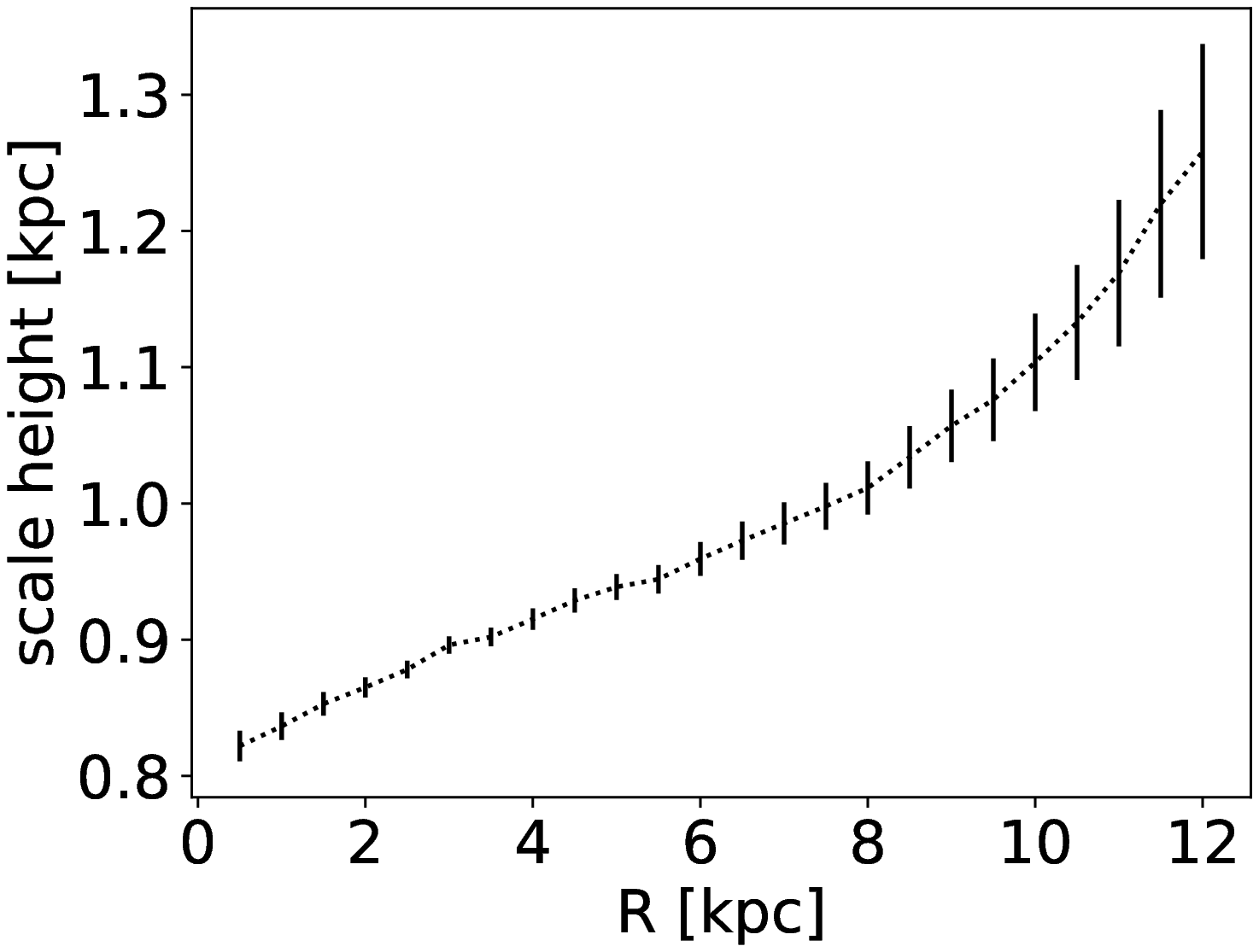}
\caption{The scale height of the EPG layer predicted by our best-fitting fountain + corona accretion model for NGC~2403.}
\label{fig:scaleheight}
\end{figure}

We determine the thickness of the EPG layer in our best-fitting model by fitting the vertical density profiles at different radii with exponential functions. Fig.~\ref{fig:scaleheight} shows the scale height of the EPG in our best-fitting fountain + corona accretion model as a function of {radius. The scale height is calculated only out} to $R=12.5$\,kpc, {as fountain clouds beyond this radius are too rare to provide a reliable vertical profile}. Overall, the thickness of the gas layer increases {slightly} with radius, which is what we would expect given that the gravitational potential is shallower {in the outer parts of the galaxy (we have assumed that $h_v$ is constant with radius for simplicity, see also Section~\ref{sec:lagexpri})}. This makes the orbits more extended in the outer region than in the inner region. The flux-weighted average scale height of our EPG model is 0.93$\pm0.003$\,kpc, compatible with the scale height derived in the kinematic model in \cite{Marasco19}. {Thus, the EPG layer of NGC~2403 is significantly thicker than its \ion{H}{i} disc, which has scale height comprised between 100 and 600\,pc \citep{Mancera22}}.

\subsubsection{EPG rotational lag}

\begin{figure}
\centering
\includegraphics[scale=0.3]{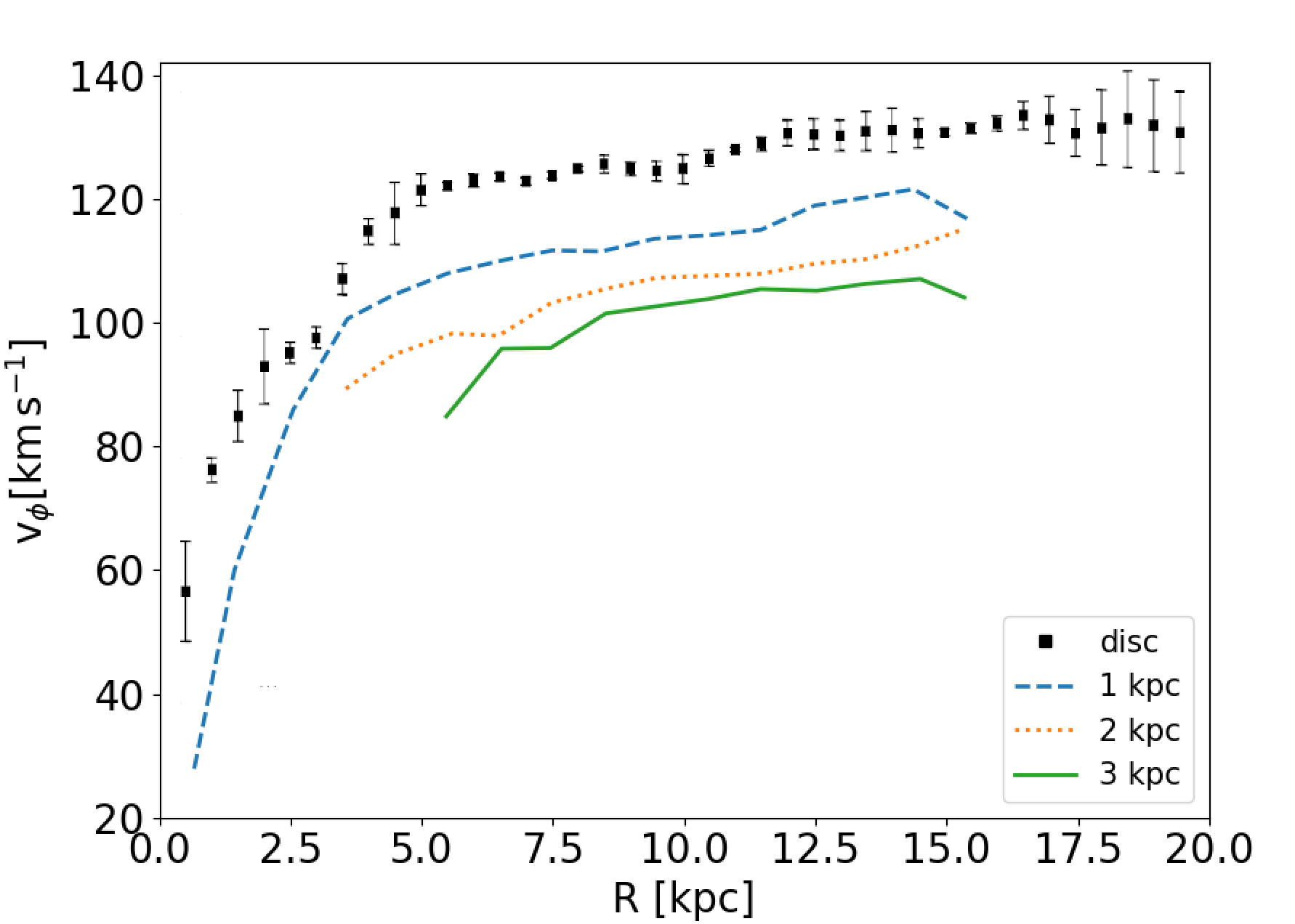}
\caption{Rotational velocities for the EPG layer at different heights from the plane (solid/dashed/dotted lines), compared to the disc rotation curve (black squares with error bars) given by \citet{Fraternali02}. Velocities are derived from our best-fitting fountain + corona accretion model by taking the flux-weighted average of azimuthal velocity $v_\phi$ at given $(R,z)$ locations.}
\label{fig:lag_rot}
\end{figure}

Fig.~\ref{fig:lag_rot} shows the rotation curves of the EPG layer at different heights above the disc. These curves are derived from our best-fitting fountain + corona accretion model by taking the flux-weighted mean value of the azimuthal velocities of the particles {in a given bin of radius and height}. We find that the rotation velocity of the EPG decreases with height. At $R=5.5$\,kpc ({the half-mass radius of the EPG in NGC~2403}), the velocity gradient is around $-10.0\pm2.7\,\mathrm{km\,s^{-1}\,kpc^{-1}}$. {This gradient is} consistent with the {velocity} gradient {of} $-11.7\pm0.5\,\mathrm{km\,s^{-1}\,kpc^{-1}}$ inferred {by} \citet{Marasco19}, {who modelled the EPG of NGC~2403 with simplified geometric and kinematic assumptions, and} therefore intrinsically differs from our dynamical model. Our results are also {comparable with the velocity gradient $-15\pm0.5\,\mathrm{km\,s^{-1}\,kpc^{-1}}$ directly measured in the edge-on galaxy NGC~891.}

\subsection{Gas flows and accretion in NGC~2403}
\label{sec:flows}
\begin{figure}
\centering
\includegraphics[scale=0.38]{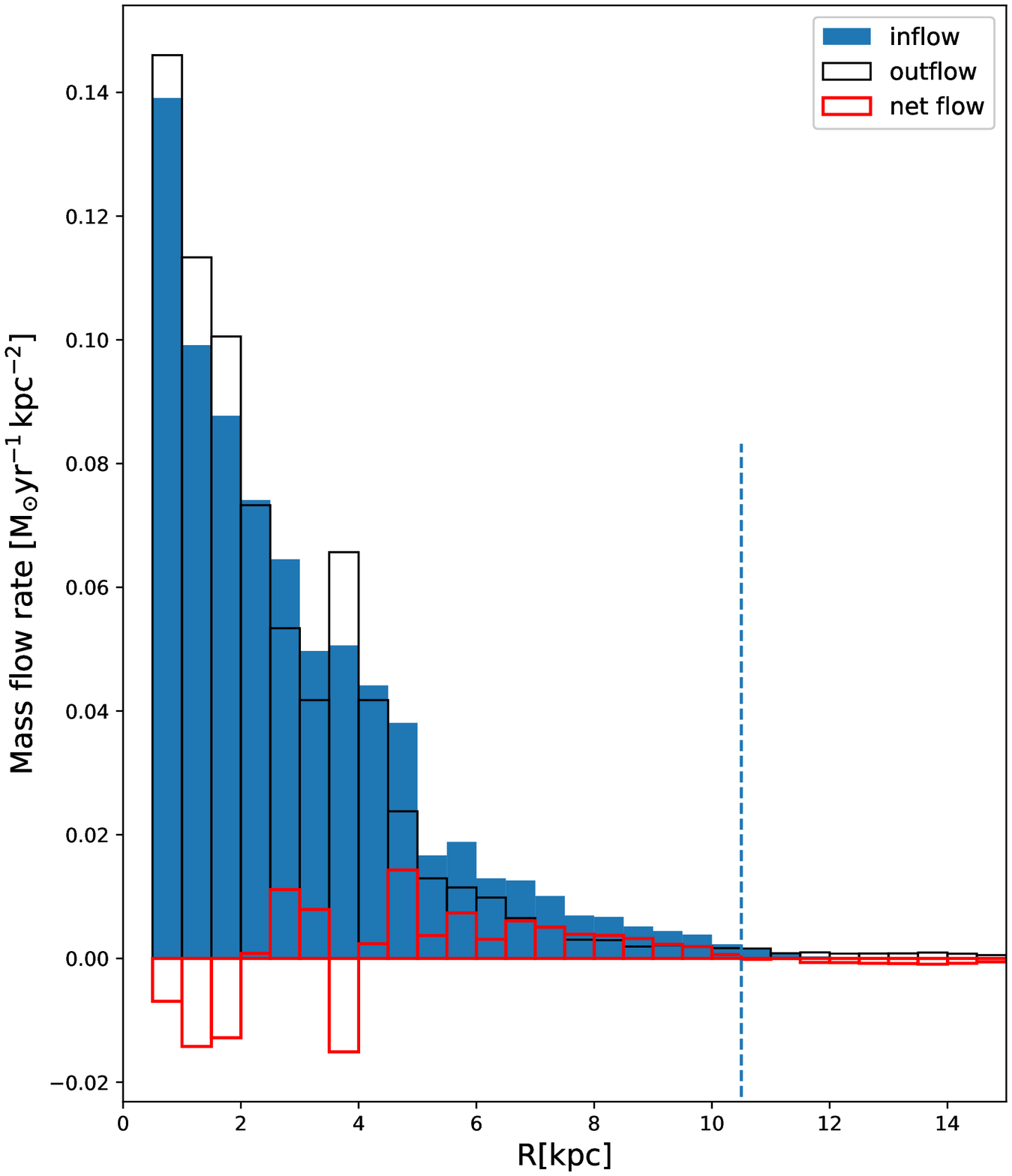}
\includegraphics[scale=0.45]{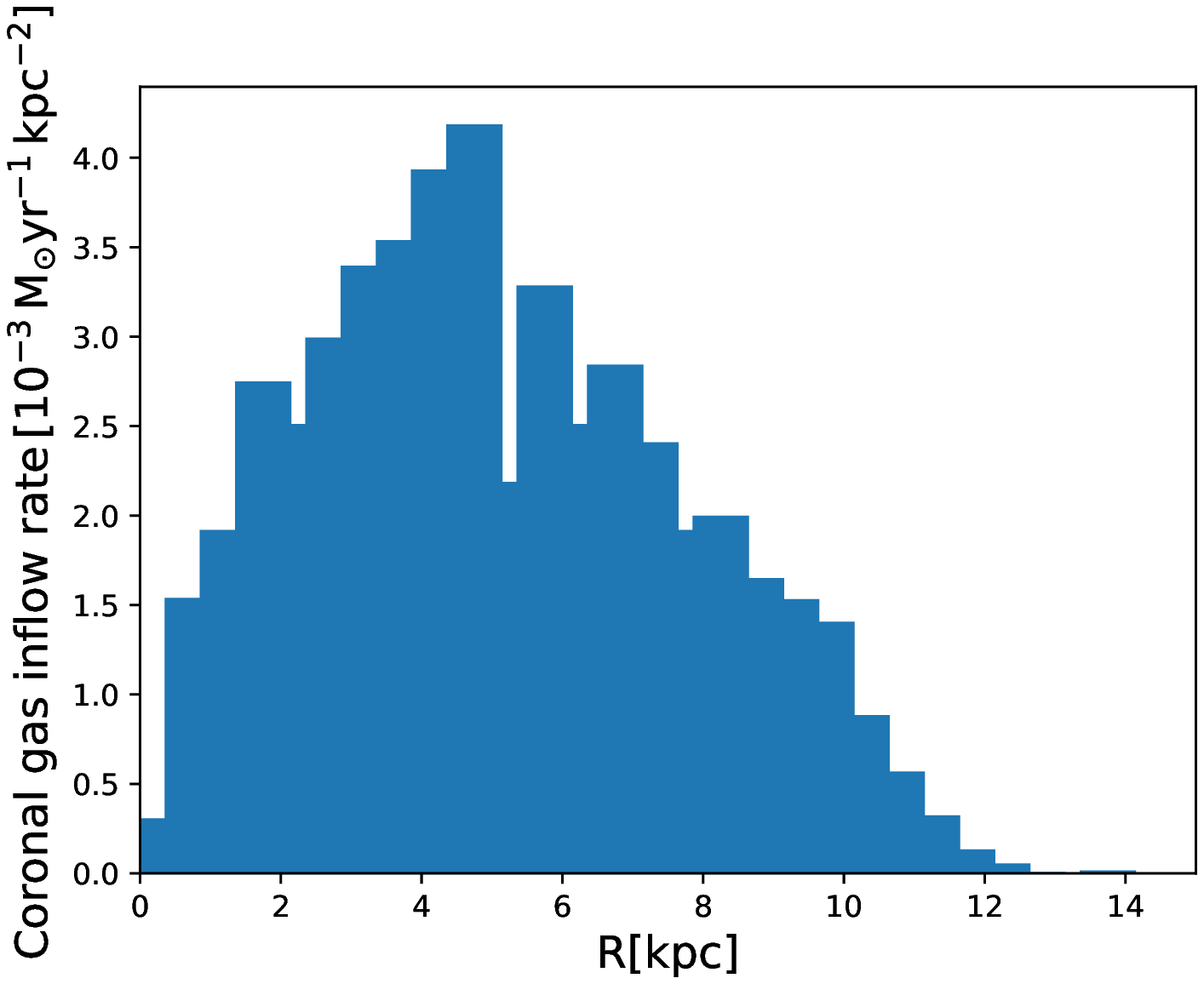}
\caption{Inflow and outflow rate surface density as a function of radius predicted by our best-fitting fountain + corona accretion model of NGC~2403. Top panel: inflow rates (blue bars), outflow rates (black bars), and net flow rates (red bars: inflow$-$outflow; positive values indicate net inflow). {The vertical dashed line at 10.5\,kpc marks the boundary where the net flow changes from inflow to outflow.} Bottom panel: inflow rate surface density contributed by corona accretion, the integration of which gives us the global accretion rate of $0.8\,\mathrm{M}_\odot\,\mathrm{yr}^{-1}$.} 
\label{fig:flow}
\end{figure}

Fig.~\ref{fig:flow} shows the inflow and outflow rates as a function of radius predicted by our best-fitting fountain + corona accretion model. {The shape of the outflow rate profile strictly follows that of the SFR profile shown in Fig.~\ref{fig:sfr}. This is true by construction, as explained in Section~\ref{sec:model}. The mass loading factor ({defined as the ratio of the mass outflow rate to the SFR and therefore is proportional to} the normalisation factor free parameter in our model) is however a prediction of our model, and we find a value of around 9.5}. The inflow rate at a given radius is given by the combination of fountain clouds and accreted coronal particles that fall onto the disc per unit time and area. Since fountain clouds do not fall back onto the disc at the same radius as they are ejected {and collect additional gas condensed from the corona as they fall}, the inflow rates do not precisely follow the outflow-rate trend but show a somewhat smoother distribution. 

We also present the net flow rate (where inflow is defined as positive value) as a function of radius in Fig.~\ref{fig:flow} top panel. The first evident feature is that the net flow is much lower than both outflow and inflow across the disc, except for the very outer parts. Also, except for some fluctuation in the innermost region (within $R=4$\,kpc), the overall tendency is net inflow in the inner region {($R<10.5$\,kpc, the vertical dashed line in Fig.~\ref{fig:flow} top panel)} and net outflow in the outer region. {The net inflow is mostly due to condensation of the hot corona, while the net outflow in the outer region can be explained by the fact that} the interaction between fountain gas and the corona results in inward orbits for the former: cloud particles are more likely to fall back to the plane at a radius smaller than their ejected radius (see Fig.~8 in \citealt{Fraternali17}). 

As we discussed in Section~\ref{sec:introduction}, accretion of the CGM onto the disc is crucial for feeding star formation and is also a key process in the evolution of a galaxy. The details of this process are however not well understood. Now with our best-fitting fountain + corona accretion model, we can predict the accretion rate as a function of radius, shown in the bottom panel of Fig.~\ref{fig:flow}. {Despite star formation being the origin of the fountain cycle, the fountain-driven accretion rate does not follow the profile of the SFR surface density (shown in Fig.~\ref{fig:sfr}) and in particular, it is more skewed towards larger radii compared with the SFR surface density profile. This is due to a number of effects, the most important of which is a radially increasing orbital time, which is in turn a consequence of a varying gravitational potential { with radius}, as also discussed in Section~\ref{sec:scaleheight}. A longer orbital time causes an increase in the total condensation along a given orbit, even with a fixed accretion efficiency per unit time (i.e. $\alpha$), as assumed in our model. The accretion profile has a well-defined peak at intermediate radii and its exact position is determined by an interplay between a radially declining SFR surface density and a radially increasing duration of the orbits {(see also M12 for the Milky Way)}.}

{The gas accretion rate that comes from corona condensation is at every radius a minor fraction of the overall gas inflow ($\sim10\%$; see Fig.~\ref{fig:flow}). Compared to the total accretion rate of $0.8\,\mathrm{M}_\odot\,\mathrm{yr}^{-1}$, the total inflow and outflow rates are $6.48\,\mathrm{M}_\odot\,\mathrm{yr}^{-1}$ and $5.69\,\mathrm{M}_\odot\,\mathrm{yr}^{-1}$, respectively.
Most of the gas inflow occurs as a consequence of the return to the disc of the gas ejected by the fountain. 
However, the fountain cycle by itself does not add any new gas to the disc and would not help to sustain the star formation. 
Instead, our model predicts that the fountain flow "captures" new gas from the corona that is then added everywhere across the disc to sustain the local star formation. 
Remarkably, the accretion rate that is needed to reproduce the seemingly independent kinematics of the EPG in NGC~2403 turns out to be {very similar} to the one needed to sustain its star formation.}

Overall, the accretion rate peaks at around 4.5\,kpc and the cumulative accretion rate reaches 50 per cent of the total accretion rate at 6.25\,kpc. {As we mentioned, this distribution is shifted outwards with respect to the} SFR surface density distribution, which peaks in the centre of NGC~2403 and reaches 50 per cent { of the} total SFR at 3.3\,kpc. {The relevance of this difference is} further discussed in Section~\ref{sec:growth}.

\section{Discussion}
\label{sec:discussion}

\subsection{Reliability of the fountain + corona accretion model}
\label{sec:lagexpri}

{In this paper, we have investigated gas accretion as the potential mechanism to maintain star formation in NGC~2403 and found a remarkable consistency between the accretion rate predicted by our model and the SFR. However, accretion is not the only fuelling mechanism. Several studies have pointed out the importance of stellar mass loss in extending gas consumption timescales \citep[e.g.][]{Sandage86,Kennicutt94} and sustaining star formation \citep[e.g.][]{Schaye10,Leitner11}. In particular, \citet[][hereafter LK11]{Leitner11} has estimated the current stellar mass loss rate of NGC~2403 to be $0.5-0.79\,\mathrm{M_\odot}\,\mathrm{yr^{-1}}$ (depending on the underlying initial mass function), which seems to eliminate the need of gas accretion. However, this mass loss rate was calculated in LK11 assuming a SFR of $1.3\,\mathrm{M_\odot}\,\mathrm{yr^{-1}}$, implying that the stellar mass loss can sustain at most 60\% of the SFR of NGC2403, while at least 40\% must be due to gas accretion. Note that the estimation of the mass loss rate is dependent on the SFR: a lower SFR would result in a lower mass loss rate (although not necessarily in proportion). Overall, we conclude that gas accretion is still necessary to sustain the SFR in NGC~2403 within the circumstances explored by the LK11 model.} 

In Section~\ref{sec:results} we explored four free parameters that are crucial for our EPG dynamical model. However, construction of the model also involves other parameters and ingredients for which we make specific choices. Below we discuss the limitations and reliability of our model.

The gravitational potential of NGC~2403 used in this paper is generated from a mass model consisting of three components: a stellar disc, a gaseous disc, and a dark matter halo. The parameters of the mass model are inferred via rotation curve decomposition (FB06). Given that the circular velocity generated from the mass model is consistent with the rotation curve of NGC~2403 (see FB06), we conclude that the gravitational potential is robust. The only uncertainty is related to the fraction of the stellar disc contribution to the potential, parametrised by the mass-to-light ratio. {The gravitational potential used in the above analysis was based on the maximum-disc model shown in Table~\ref{tab:sffr}. It is however noteworthy that the minimum disc potential in FB06 is in fair agreement with those derived more recently with more sophisticated methods \citep{Mancera22}.} FB06 have experimented with both maximum disc and minimum disc potentials and showed that the dynamics of the EPG does not change significantly.

An assumption of our model is the existence of a uniform characteristic outflow velocity at all radii, whereas the varying stellar feedback activities might lead to outflow velocities changing with radius. Allowing spatial variations in the characteristic outflow velocity is a potential improvement for this kind of study. This has been briefly explored in FB06 to generate specific features in N2403 {(e.g. the filament shown in channel 104.1$\,\mathrm{km\,s^{-1}}$ and channel 135.0$\,\mathrm{km\,s^{-1}}$ of Fig.~14 in FB06)} that are otherwise not reproduced. However, {exploring the variation of $h_v$ with radius} would introduce at least one extra free parameter, which would significantly complicate our exploration of the parameter space. {Overall, the global kinematics of the EPG in NGC~2403 appears to be well reproduced by a constant characteristic outflow speed across the disc.}

{In the fountain + corona accretion scenario,} the acceleration of fountain gas is directly dictated, {besides by gravity}, by the velocity difference between the fountain and the corona. In our model, we assume a relative azimuthal velocity of $75\,\mathrm{km\,s^{-1}}$ between the fountain gas and the corona, based on hydrodynamical simulations \citep{Marinacci11}. {Such a high relative velocity would imply a rather slowly rotating corona in NGC~2403, given the disc rotation of around $130\,\mathrm{km\,s^{-1}}$ (FB06)}. We have therefore tested models with a lower relative velocity of  $45\,\mathrm{km\,s^{-1}}$ that result in { nearly identical} best-fitting parameters as in Section~\ref{sec:results_n2403} { except for} a higher condensation rate (4.2$\pm1.2$\,Gyr$^{-1}$), which corresponds to a global accretion rate of $1.1^{+0.3}_{-0.2}\,\mathrm{M}_\odot\,\mathrm{yr}^{-1}$ {(the best-fitting results are listed in Table~\ref{tab:bestfit})}. {This higher rate is not surprising. In our model, as a consequence of condensation, the coronal gas joins the cold/warm phase of the fountain gas such that the velocity of a single cloud evolves as a combination (mass-weighted average) of the kinematics of the two components (cloud and condensed material). If the velocity difference between these two components is reduced, one needs a larger accretion rate (more condensed material) to produce the same effect in the combined kinematics.} It is noteworthy that EPG models built with a lower relative velocity have lower velocity gradients than what we show in Fig.~\ref{fig:lag_rot}. However, the difference ($1.0\,\mathrm{km\,s^{-1}\,kpc^{-1}}$) is negligible, given that the uncertainty for our measurement is  $2.7\,\mathrm{km\,s^{-1}\,kpc^{-1}}$. 

The separation of EPG emission from the datacube { is an important ingredient of our method}. The reliability of our strategy for masking the disc emission has been verified in several previous studies \citep[e.g.][]{Fraternali02,Marasco19,Li21}. {We have tested the robustness of our results by fitting the data without masking the peculiar \ion{H}{i} filament of NGC~2403, finding the same normalisation factor} as shown in Table~\ref{tab:bestfit}, but an $h_v$ of $60\,\mathrm{km\,s^{-1}}$, an $f_{\mathrm{ion}}$ of 0, a condensation rate of 4.8\,Gyr$^{-1}$, leading to an accretion rate of $1.28\,\mathrm{M}_\odot\,\mathrm{yr}^{-1}$ {(all parameters are compatible with those of our fiducial model within the errors.)}. Thus models with slightly higher outflow velocities and condensation rates are preferred to account for the filament in NGC~2403, but the overall validity of our results is not particularly affected by our masking.

In conclusion, the construction of our dynamical model is robust. The variation of certain ingredients leads to small changes in the model best-fitting parameters but does not alter our main conclusion: the EPG of NGC~2403 is produced by a combination of galactic fountain clouds and gas accretion from the {condensation of the hot CGM at a rate compatible with the SFR of the galaxy}.

\subsection{Can the fountain + corona accretion sustain the inside-out growth of the disc?}
\label{sec:growth}
{Since accretion is a {key} source to fuel further star formation,} the outward shift of the accretion (compared to the SFR) shown in Section~\ref{sec:flows} suggests a potential inside-out redistribution of gas and star formation activities in the future, which has been predicted by cosmological simulations (e.g. \citealt{Grand17}) and supported by many observations (e.g. \citealt{Wang11,vanderwel14,Pezzulli15}). \citet{Pezzulli15} also provided measurements of the specific radial growth rate, $\nu_R\equiv(1/R_*)\times\mathrm{d}R_*/\mathrm{d}t$, where $R_*$ is the scale length of the stellar disc, for a sample of galaxies including NGC~2403. {Furthermore, a cosmological/zoom-in simulation \citep{Grand19} also found that fountain clouds can acquire angular momentum via interaction with the CGM.}

{To verify whether the gas accretion due to a galactic fountain can be deemed responsible for this growth}, we calculated the {variation in time of the }specific angular momentum ${\mathrm{d}j}/{\mathrm{d}t}$ of the stellar disc \citep[a direct tracer of disc growth;][]{Mo98,Posti19b} due to accretion, {under the simplifying assumption that the next generation of stars will be formed out of the newly accreted gas. This gives}
\begin{eqnarray}
\frac{\mathrm{d}j}{\mathrm{d}t}&=&\frac{\mathrm{d}(J/M)}{\mathrm{d}t} \nonumber \\
&=&\frac{1}{M}\frac{\mathrm{d}J}{\mathrm{d}t}-\frac{J}{M^2}\frac{\mathrm{d}M}{\mathrm{d}t},
\end{eqnarray}
where $J$ and $M$ ($7.2\times10^{9}\,\mathrm{M_{\odot}}$) are the angular momentum and mass of the stellar disc. We estimate $J$  as $J=2MV_\mathrm{flat}R_*$ \citep{Romanowsky12}, where $V_\mathrm{flat}$ is the rotational velocity of the flat part of the rotation curve ($130\mathrm{\,km\,s^{-1}}$) and ${R_*}$\,=\,2.0\,kpc \citep[values from][]{Fraternali02}. The time derivative of the angular momentum $\mathrm{d}J/\mathrm{d}t$ is given by
\begin{eqnarray}
\frac{\mathrm{d}J}{\mathrm{d}t}&=&\frac{\mathrm{d}J_\mathrm{in}}{\mathrm{d}t}-\frac{\mathrm{d}J_\mathrm{out}}{\mathrm{d}t} \nonumber\\
&=&2\pi\int_{0}^{\mathrm{R}} R'^2\mathrm{\mathcal{F}_\mathrm{in}}(R') \overline{V_\mathrm{in}(R')}\,\mathrm{d}R' \nonumber\\
&&-2\pi\int_{0}^{\mathrm{R}} R'^2\mathrm{\mathcal{F}_\mathrm{out}}(R') \overline{V_\mathrm{out}(R')}\,\mathrm{d}R',
\end{eqnarray}
where {$\mathrm{\mathcal{F}_\mathrm{in}}$ ($\mathrm{\mathcal{F}_\mathrm{out}}$)} is the inflow (outflow) surface density rate given in Section~\ref{sec:flows}, $\overline{V_\mathrm{in}(R')}$ ($\overline{V_\mathrm{out}(R')}$)  is the average rotational velocity of all cloud particles falling onto (ejected from) the disc at radius $R'$, obtained from our model by tracking the outflow and inflow radius and velocity of all fountain clouds. The time derivative of the mass, $\mathrm{d}M/\mathrm{d}t$, is by definition the accretion rate of new gas given by the model.

Implementing the above equation to our best-fitting model, we have $\mathrm{d}j/\mathrm{d}t=-2.6\times10^{-8}\,\mathrm{km\,s^{-1}\,kpc\,yr^{-1}}$. {This would indicate that the gas accreted through the fountain cannot be solely responsible for the observed inside-out growth of the disc. Part of this growth should then be ascribed to gas that is already present in the disc. This is a viable option, as the gas in the disc is known to be located, on average, at larger radii compared to the stellar component \citep[e.g.][]{Fraternali02}. This solution is, however, only partly satisfactory, as the gas reservoir at these large radii would, without replacement, be consumed on a relatively short timescale \citep[a few Gyr; see e.g.][]{Fraternali12}, implying that the growth of the disc would not be sustainable in the long term. 

With these considerations in mind, we stress that our calculation of $\mathrm{d}j/\mathrm{d}t$, presented above, very much depends on the value that we are assuming }{for the rotational speed of the corona, which is, as we discussed above, very uncertain.} Interestingly, when assuming the rotational lag between the fountain and the hot gas is $45\mathrm{\,km\,s^{-1}}$ { (the third model in Table~\ref{tab:bestfit})}, we have $\mathrm{d}j/\mathrm{d}t=1.5\times10^{-8}\,\mathrm{km\,s^{-1}\,kpc\,yr^{-1}}$, {which indicates an inside-out growth. {Combining the current value of the specific angular momentum $j$ and its derivative $\mathrm{d}j/\mathrm{d}t$, we can easily derive the specific angular momentum growth rate, which we define \citep[following][]{Pezzulli15} as $\nu_j\,\equiv\,(1/j)\times\mathrm{d}j/\mathrm{d}t$. We find a value of $\nu_j = 2.88\,\times\,10^{-2} \, \textrm{Gyr}^{-1}$, in excellent agreement with the specific radial growth rate $\nu_R\,=\,(2.93\pm0.16)\,\times\,10^{-2} \; \textrm{Gyr}^{-1}$ measured by \citet{Pezzulli15} for NGC~2403. The two quantities $\nu_j$ and $\nu_\textrm{R}$ are comparable and are in fact expected to be equal, as long as the rotation curve of the galaxy can be considered approximately stationary with time\footnote{This is immediately seen by taking the time derivative of the equation $j=2V_\mathrm{flat}R_*$.}.
We have therefore found that our model with a reduced rotational lag is in remarkable quantitative agreement with the galactic fountain being the main source of the observed inside-out growth in NGC 2403.}

It is important to note that in the absence of triggered condensation, a galactic corona will be expected to cool in the very inner parts, where its density tends to be higher, thus producing the accretion of low angular momentum gas that then would need to be expelled via strong feedback \citep[e.g.][]{Brook12}. Instead, when the cooling is triggered by the fountain, the location of the bulk of the gas accretion is naturally shifted to outer radii for the reasons described in Section~\ref{sec:flows}. This phenomenon had been indicated as plausibly compatible with the inside-out growth of discs \citep{Pezzulli16}, but this is the first time that quantitative evidence is provided.}

\section{Conclusion}
\label{sec:conclusion}
{In this work, we have modelled the distribution and kinematics of the neutral extra-planar gas (EPG) in the late-type nearby galaxy NGC~2403 using a dynamical model of galactic fountain.} In this model, stellar feedback activities continuously eject gas from the galaxy disc, which travels through the halo and falls back to the disc. This gas cycle brings metal-rich and cold/warm gas to mix and interact with the hot corona, significantly reducing its cooling time, and leading to condensation and accretion of some coronal gas onto the disc. Due to angular momentum exchange between the fountain clouds and the corona, this interaction is expected to leave a signature in the kinematics of the \ion{H}{i} gas at the disc--halo interface. The application of our models to the data leverage this signature to infer, along with other parameters, the efficiency of the condensation process and the accretion rate of coronal gas onto the disc.

While these models have been applied extensively to the EPG of the Milky Way \citep{Marasco12,Marasco13,Fraternali13,Fraternali15}, so far applications to external galaxies were limited to the preliminary studies of FB06 and FB08, {which did not include a rotating corona nor a statistically meaningful exploration of the parameter space.} This study presents the first detailed application of the current fountain accretion framework to an external galaxy.  Our results are summarised as follows:
\begin{enumerate}
\item {The galactic fountain framework can reproduce most of the neutral EPG features in NGC~2403. A model where the fountain clouds interact with the hot corona is statistically preferred compared to a pure fountain model without interaction with the hot CGM.}

\item The best-fitting model requires a fountain with a characteristic outflow velocity of $50\pm10\,\mathrm{km\,s^{-1}}$, with the gas being ionised {for some time after ejection and then recombining}. Recombination {appears to occur on average }when its vertical velocity has been reduced by about 40 per cent. 

\item The \ion{H}{i} EPG in NGC~2403 inferred from the best-fitting model has a total EPG mass of $4.7^{+1.2}_{-0.9}\times 10^8\,\mathrm{M_\odot}$, with an average scale height of $0.93\pm0.003\,$kpc and a vertical gradient in rotational velocity of $-10.0\pm2.7\,\mathrm{km\,s^{-1}\,kpc^{-1}}$. Our values are compatible with a previous estimate of \citet{Marasco19}, which was derived with {simpler phenomenological} approaches.

\item {Our model predicts a condensation rate of 2.4\,Gyr$^{-1}$ (4.2\,Gyr$^{-1}$ ) for the hot CGM, leading to a total accretion rate of $0.8\,\mathrm{M}_\odot\,\mathrm{yr}^{-1}$ ($1.1\,\mathrm{M}_\odot\,\mathrm{yr}^{-1}$) when assuming the rotational lag between the fountain and the hot gas is $75\mathrm{\,km\,s^{-1}}$ ($45\mathrm{\,km\,s^{-1}}$)}, similar to the star formation rate $0.6\,\mathrm{M}_\odot\,\mathrm{yr}^{-1}$ of NGC~2403, suggesting corona accretion as a viable mechanism to maintain the star-formation rate in this galaxy.

\item The accretion rate surface density profile predicted by our model is radially more extended than the
star-formation-rate surface density. {We have also shown that, if the rotation velocity of the corona is larger than a certain threshold, the specific angular momentum growth rate predicted by our model is in excellent agreement with the observed inside-out growth rate in NGC~2403.} The fountain-driven
accretion process can therefore be responsible for the inside-out
growth of its stellar disc.

\end{enumerate}

\section*{Acknowledgements}
The authors would like to thank {an anonymous referee for helpful comments and } Cecilia Bacchini for collecting and providing the \ion{H}{i}, H$_2$, and star-formation-rate data of NGC~2403. 
AL was supported by the Netherlands Research School for Astronomy (Nederlandse Onderzoekschool voor Astronomie, NOVA), Network 1, Project 10.1.5.9 WEAVE.
{GP acknowledges support from the Netherlands Research School for Astronomy (Nederlandse Onderzoekschool voor Astronomie, NOVA) through project 10.1.5.18.}

\section*{Data Availability}
The data underlying this article were obtained by \citet{Fraternali02} with the CS configuration
of the VLA and were later included in the HALOGAS survey, which is available at https://www.astron.nl/halogas.
 
\bibliographystyle{mnras}
\bibliography{references}

\appendix
\renewcommand\thefigure{\Alph{section}\arabic{figure}} 
\section{2D marginalised posterior probability distribution}
\label{appen:a}
\begin{figure*}
\centering
\includegraphics[scale=0.31]{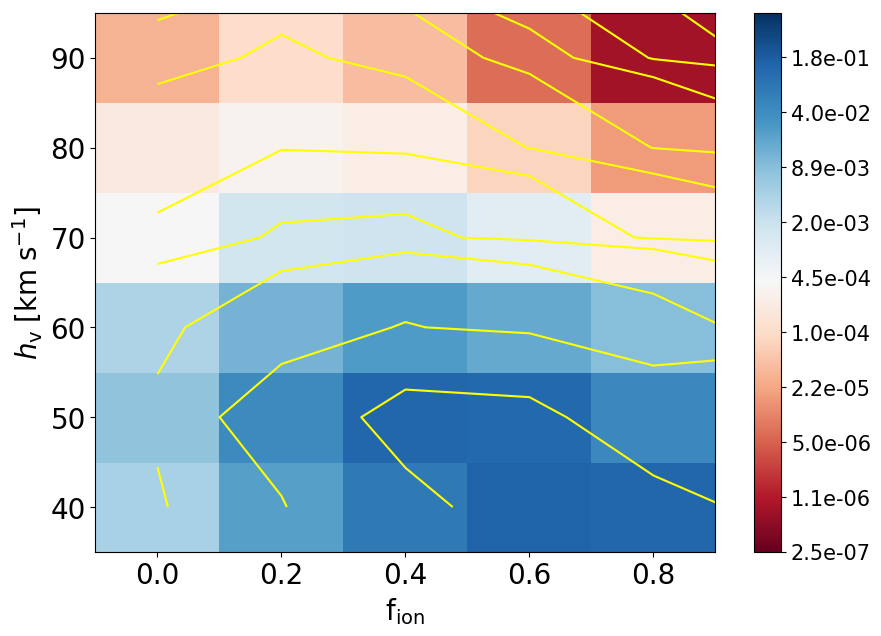}
\includegraphics[scale=0.31]{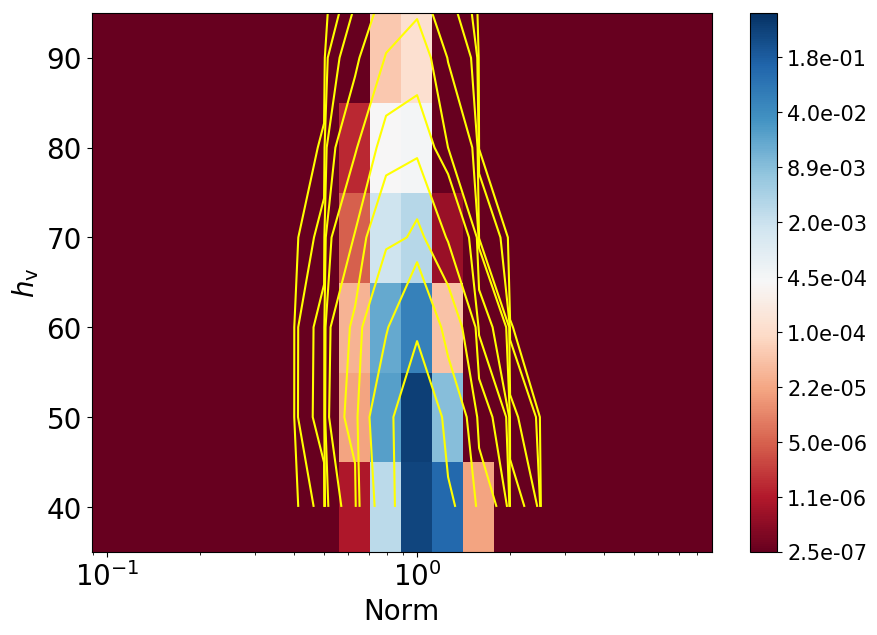}
\includegraphics[scale=0.31]{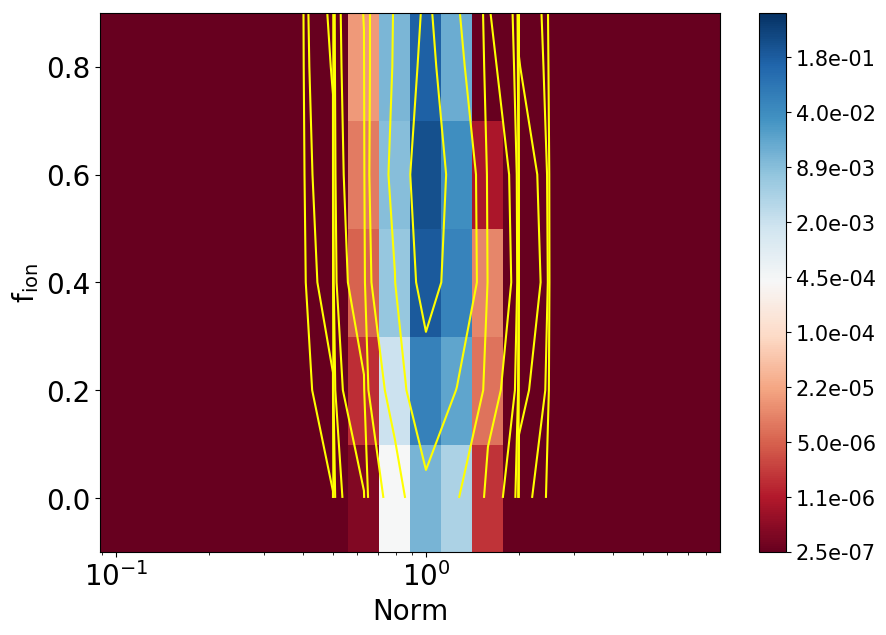}

\caption{2D marginalised posterior probability distribution for our pure fountain models onto different 2D spaces: upper left -- ($h_v$,$f_\mathrm{ion}$), upper right -- ($h_v$,Norm), lower-left -- ($f_\mathrm{ion}$,Norm).  Iso-probability contours (in yellow) correspond to 2.51e-07, 1.06e-06, 4.47e-06, 1.88e-05, 7.94e-05, 3.34e-04, 1.41e-03, 5.96e-03,2.51e-02, 1.05e-01.}
\label{fig:contn2403gf}
\end{figure*}

\begin{figure*}
\centering
\includegraphics[scale=0.31]{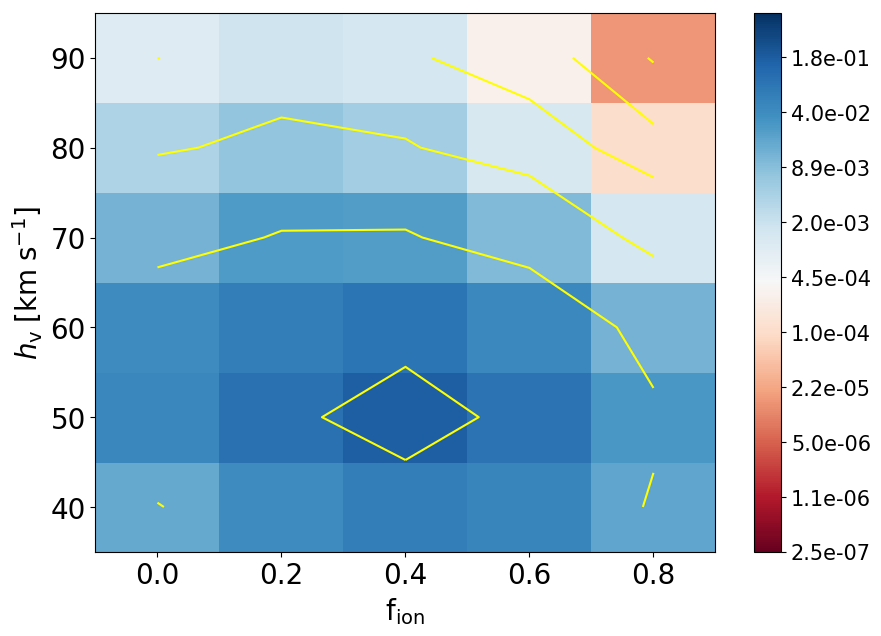}
\includegraphics[scale=0.31]{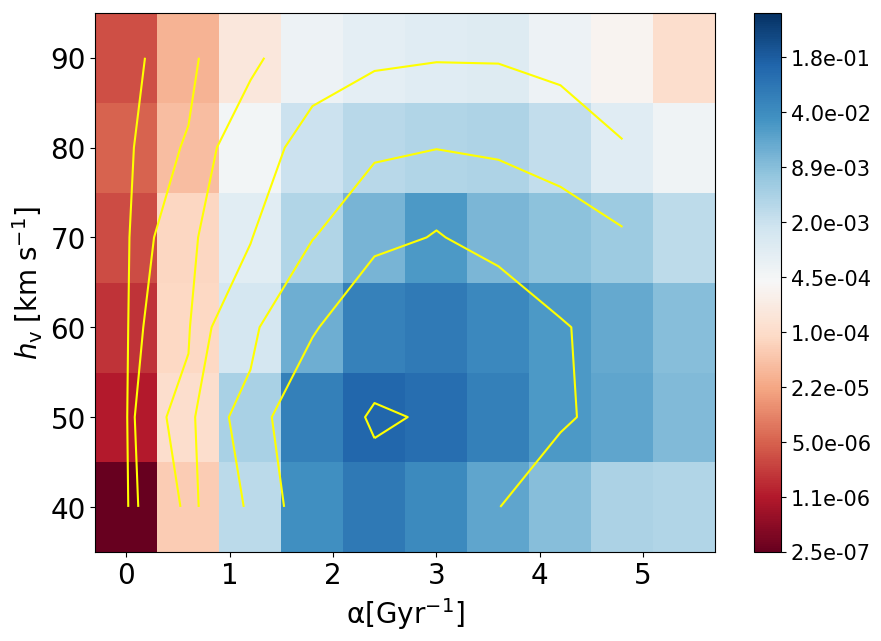}
\includegraphics[scale=0.31]{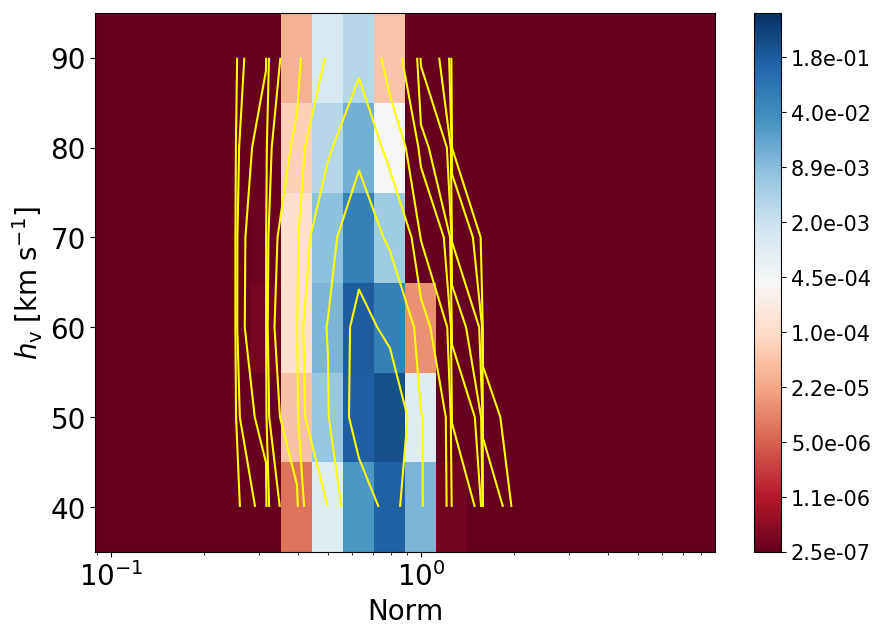}
\includegraphics[scale=0.31]{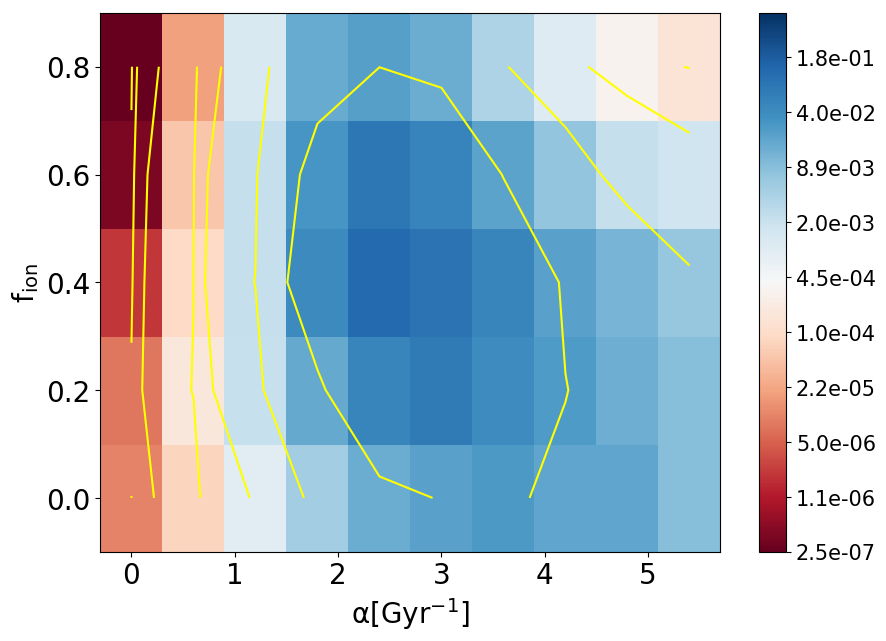}
\includegraphics[scale=0.31]{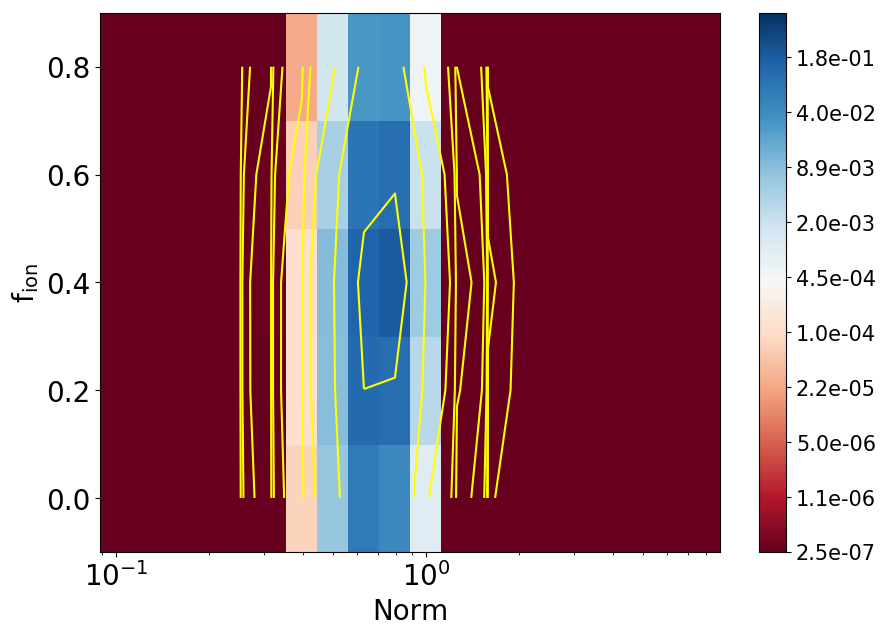}
\includegraphics[scale=0.31]{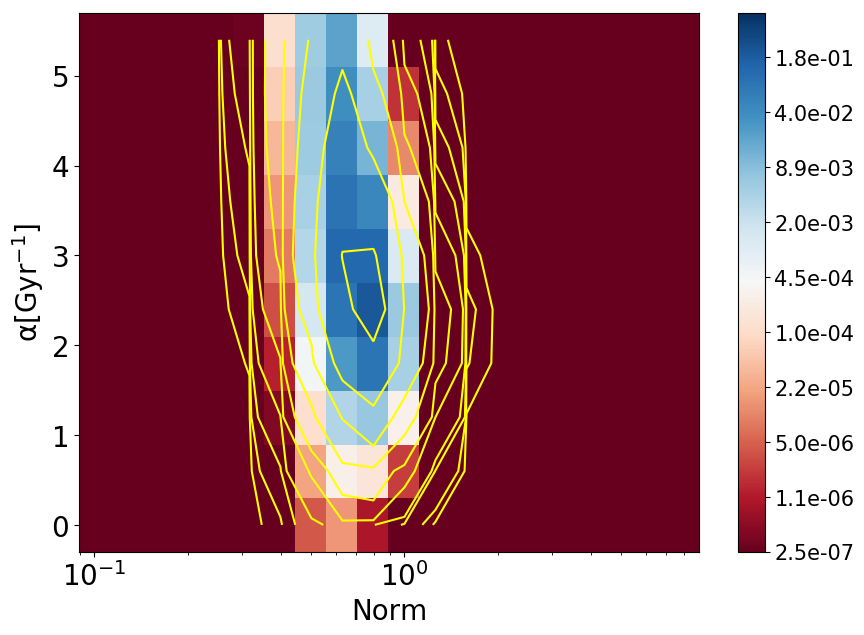}
\caption{2D marginalised posterior probability distribution for our fountain + corona accretion models onto different 2D spaces: upper-left -- ($h_v$,$f_\mathrm{ion}$), upper-right -- ($h_v$,$\alpha$), middle-left -- ($h_v$,Norm), middle-right -- ($f_\mathrm{ion}$,$\alpha$), lower-left -- ($f_\mathrm{ion}$,Norm), lower-right -- ($\alpha$,Norm).  Iso-probability contours (in yellow) correspond to 2.51e-07, 1.06e-06, 4.47e-06, 1.88e-05, 7.94e-05, 3.34e-04, 1.41e-03, 5.96e-03,2.51e-02, 1.05e-01.}
\label{fig:contn2403ac}
\end{figure*}

2D marginalised posterior probability distribution maps and contours of the three(four)-dimensional grids of free parameters {(summarised in Table~\ref{tab:free_param})} for pure fountain (fountain + corona accretion) models, shown in  Figs.~\ref{fig:contn2403gf} and \ref{fig:contn2403ac} respectively.

\label{lastpage}

\clearpage
\end{document}